\documentclass[superscriptaddress,nofootinbib]{revtex4}

\usepackage[utf8]{inputenc}
\usepackage{appendix,typearea,amsmath,color,url,epsfig,graphicx,amsthm,ulem,xcolor}
\newcommand{\bold}[1]{{\bf #1}}

\newcommand{\fgoal}[0]{f_{\rm goal}}
\newcommand{\uenc}[0]{U_{\rm enc}}
\newcommand{\boldx}[0]{\bold{x}}
\newcommand{\fq}[0]{f^{\rm Q}}
\newcommand{\fqarray}[0]{(\fq\xa_1, \fq\xa_2,\cdots, \fq\xa_{n_0})}
\newcommand{\boldfq}[0]{\bold{f}^Q\xa}
\newcommand{\boldfqx}[1]{\bold{f}^Q({#1})}
\newcommand{\thetatau}{\boldsymbol{\theta}(\tau)}
\newcommand{\thetat}{\boldsymbol{\theta}(t)}
\newcommand{\inttwodesign}[1]{\int_{\rm 2-design}d#1}
\newcommand{\thetazero}{\boldsymbol{\theta}(0)}
\newcommand{\fc}[0]{f^{\rm C}_{\thetat}}
\newcommand{\xa}[0]{(\boldx^a)}
\newcommand{\xb}[0]{(\boldx^b)}
\newcommand{\rhok}[1]{\rho^k_{\bold{#1}}}
\newcommand{\uencxa}[0]{\uenc \xa}

\newcommand{\boldalpha}[2]{\boldsymbol{\alpha}^{(#1)}(\bold{x}^{#2})}
\newcommand{\tildealpha}[2]{\tilde{\boldsymbol{\alpha}}^{(#1)}(\bold{x}^{#2})}
\newcommand{\cov}[1]{\boldsymbol{\Sigma}^{(#1)}(\bold{x}, \bold{x}^{\prime})}
\newcommand{\covdot}[1]{\Dot{\boldsymbol{\Sigma}}^{(#1)}(\bold{x}, \bold{x}^{\prime})}
\newcommand{\covx}[3]{\boldsymbol{\Sigma}^{(#1)}(#2, #3)}
\newcommand{\covq}[1]{\boldsymbol{\Sigma}_Q^{(#1)}(\bold{x}, \bold{x}^{\prime})}
\newcommand{\covqx}[3]{\boldsymbol{\Sigma}_Q^{(#1)}(#2, #3)}
\newcommand{\kernel}[1]{K^{(#1)}(\bold{x}, \bold{x}^{\prime}, t)}
\newcommand{\kernelx}[3]{K^{(#1)}(#2, #3, t)}

\newcommand{\ntk}[1]{\boldsymbol{\Theta}^{(#1)}(\bold{x}, \bold{x}^{\prime})}
\newcommand{\ntkx}[3]{\boldsymbol{\Theta}^{(#1)}(#2, #3)}
\newcommand{\ntkq}[1]{\boldsymbol{\Theta}_Q^{(#1)}(\bold{x}, \bold{x}^{\prime})}
\newcommand{\ntkqx}[3]{\boldsymbol{\Theta}_Q^{(#1)}(#2, #3)}
\newcommand{\ft}[0]{f_{\thetat}}
\newcommand{\ftau}[0]{f_{\thetatau}}
\newcommand{\fzero}[0]{f_{\thetazero}}
\newtheorem{th.}{Theorem}
\newtheorem{co.}{Corollary}

\newtheorem{innercustomthm}{Theorem}
\newtheorem{lemma}{Lemma}
\newenvironment{customthm}[1]
  {\renewcommand\theinnercustomthm{#1}\innercustomthm}
  {\endinnercustomthm}

\makeatletter
\newcommand{\figcaption}[1]{\def\@captype{figure}\caption{#1}}
\newcommand{\tblcaption}[1]{\def\@captype{table}\caption{#1}}
\makeatother
\usepackage{threeparttable} %

\usepackage{subcaption}
\usepackage{booktabs}
\usepackage{multirow}
\usepackage{comment}

\typearea{14}
\begin{document}
\title{
Quantum-classical hybrid neural networks in the neural tangent kernel regime}

\author{Kouhei~Nakaji}
\affiliation{Quantum Computing Center, Keio University, Hiyoshi 3-14-1, Kohoku-ku, Yokohama 223-8522, Japan}
\affiliation{Graduate School of Science and Technology, Keio University, 
Hiyoshi 3-14-1, Kohoku-ku, Yokohama, 223-8522, Japan}

\author{Hiroyuki~Tezuka}
\affiliation{
Advanced Research Laboratory, Technology Infrastructure Center,
Technology Platform,
Sony Group Corporation, 1-7-1 Konan, Minato-ku, Tokyo, 108-0075, Japan}
\affiliation{Quantum Computing Center, Keio University, Hiyoshi 3-14-1, Kohoku-ku, Yokohama 223-8522, Japan}
\affiliation{Graduate School of Science and Technology, Keio University, 
Hiyoshi 3-14-1, Kohoku-ku, Yokohama, 223-8522, Japan}

\author{Naoki~Yamamoto}
\affiliation{Quantum Computing Center, Keio University, Hiyoshi 3-14-1, Kohoku-ku, Yokohama 223-8522, Japan}
\affiliation{Department of Applied Physics and Physico-Informatics, 
Keio University, Hiyoshi 3-14-1, Kohoku-ku, Yokohama, 223-8522, Japan}

\begin{abstract}
Recently, quantum neural networks or quantum-classical neural networks (qcNN) 
have been actively studied, as a possible alternative to the conventional 
classical neural network (cNN), but their practical and theoretically-guaranteed 
performance is still to be investigated. 
In contrast, cNNs and especially deep cNNs, 
have acquired several solid theoretical basis; one of those basis is the neural 
tangent kernel (NTK) theory, which can successfully explain the mechanism of 
various desirable properties of cNNs, particularly the global convergence in the 
training process. 
In this paper, we study a class of qcNN composed of a quantum data-encoder 
followed by a cNN. 
The quantum part is randomly initialized according to 
unitary 2-designs, which is an effective feature extraction process 
for quantum states, and the classical part is also randomly initialized according 
to Gaussian distributions;
then, in the NTK regime where the number of nodes of the cNN becomes infinitely 
large, the output of the entire qcNN becomes a nonlinear function of the so-called 
projected quantum kernel. 
That is, the NTK theory is used to construct an effective quantum kernel, which is 
in general nontrivial to design. 
Moreover, NTK defined for the qcNN is identical to the covariance matrix of 
a Gaussian process, which allows us to analytically study the learning process. 
These properties are investigated in thorough numerical experiments; 
particularly, we demonstrate that the qcNN shows a clear advantage over fully 
classical NNs and qNNs for the problem of learning the quantum data-generating 
process. 
\end{abstract}

\maketitle
\def\thefootnote{}
\footnotetext{Kouhei Nakaji: These authors contributed equally to this work}
\footnotetext{Hiroyuki Tezuka: These authors contributed equally to this work}

\section{Introduction}
\label{section:introduction}

\subsection{Background -- Quantum/classical neural networks and classical neural tangent kernel}

Quantum neural networks (qNNs) or quantum classical hybrid neural networks (qcNNs) 
are systems that, based on their rich expressibility in the functional space, 
have potential of offering a higher-performance solution in various problems 
over classical means \cite{altaisky2001quantum,farhi2018classification,mitarai2018quantum,schuld2020circuit,romero2017quantum,wan2017quantum,verdon2018universal,du2020expressive,romero2021variational,beer2020training,abbas2021power,cong2019quantum,nakaji2021quantum,xia2020hybrid,mari2020transfer}. 
However, there remain two essential issues to be resolved. 
First, the existing qNN and qcNN models have no theoretical guarantee in their 
training process to converge to the optimal or even a ``good" solution. 
The vanishing gradient (or the barren plateau) issue, stating that the gradient 
vector decays exponentially fast with respect to the number of qubits, is 
particularly serious \cite{McClean2018BarrenPI}; several proposals to 
mitigate this issue have been proposed \cite{sack2022avoiding, grant2019initialization, skolik2021layerwise, dborin2022matrix, holmes2022connecting, cerezo2021cost, marrero2021entanglement, wang2021noise, patti2021entanglement, holmes2021barren}, but these are not general solutions. 
Secondly, despite the potential advantage of the quantum models in their 
expressibility, they are not guaranteed to offer a better solution over classical 
means, especially the classical neural networks (cNNs). 
Regarding this point, the recent study \cite{huang2021power} has derived a 
condition for the quantum kernel method to presumably outperform a class 
of classical means and provided the idea using the projected quantum kernel 
to satisfy this advantageous condition. 
Note that the quantum kernel has been thoroughly 
investigated in several theoretical and experimental settings 
\cite{NEURIPS2021_69adc1e1,liu2021rigorous,peters2021machine,sancho2022quantum,PRXQuantum.3.030101,schuld2021supervised,gentinetta2022complexity,thanasilp2022exponential}.
However, designing an effective quantum kernel (including the projected quantum 
kernel) is a highly nontrivial task; also, the kernel method generally requires 
the computational complexity of $O(N_D^2)$ with $N_D$ the number of data, whereas 
the cNN needs only $O(N_D)$ as long as the 
computational cost of training does not scale with $N_D$. 
Therefore, it is desirable if we could have an easy-trainable qNN or qcNN to which 
the above-mentioned advantage of quantum kernel method are incorporated.

On the other hand, in the classical regime, the neural tangent kernel (NTK)  
\cite{NEURIPS2018_5a4be1fa} offers useful approaches to analyze several fundamental 
properties of cNNs and especially deep cNNs, including the convergence properties 
in the training process. 
The NTK is a time-varying nonlinear function that appears in the dynamical 
equation of the output function of cNN in the training process. 
Surprisingly, NTK becomes time-invariant in the so-called NTK regime where the number 
of nodes of CNN becomes infinitely large; further, it becomes positive-definite via 
random initialization of the parameters. 
As a result, particularly when the problem is the least square regression, the 
training process is described by a linear differential (or difference) equation, 
and the analysis of the training process boils down to that of 
the spectra of this time-invariant positive-definite matrix. 
The literature studies on NTK that are related to our work are as follows; the 
relation to Gaussian process \cite{lee2017deep}, relation between 
the spectra of NTK and the convergence property of 
cNN \cite{arora2019fine}, and the NTK in the case of classification problem 
\cite{NEURIPS2019_62dad6e2,cao2020generalization,nitanda2019gradient,LeeXSBNSP19}.

\subsection{Our contribution}

In this paper, we study a class of qcNN that can be directly 
analyzed in the NTK regime. 
In this proposed qcNN scheme, the classical data is first encoded into the state 
of a quantum system and then re-transformed to a classical data by some 
appropriate random measurement, which can thus be regarded as a feature extraction 
process in the high-dimensional quantum Hilbert space. 
We then input the reconstructed classical data vector into a subsequent cNN. 
Finally, a cost function is evaluated using the output of cNN, and the parameters 
contained in the cNN part are updated to lower the cost. 
Note that, hence, the quantum part is fixed, implying that the vanishing gradient 
issue does not occur in our framework. 
The following is the list of results. 
\begin{itemize}
\item
The output of qcNN becomes a Gaussian process in the infinite width limit of 
the cNN part while the width of the quantum part 
is fixed, where the unitary gate determining the quantum measurement and the 
weighting parameters of cNN are randomly chosen from 
unitary 2-designs and Gaussian distributions, respectively. 
The covariance matrix of this Gaussian process is given by a function of 
projected quantum kernels mentioned in the first paragraph. 
That is, our qcNN certainly exploits the quantum feature space. 

\item
In the infinite width limit of cNN, the training dynamics in the functional space 
is governed by a linear differential equation characterized by the corresponding 
NTK, meaning the exponentially-fast convergence to the global solution if NTK is 
positive-definite; a condition to guarantee the positive-definiteness is also 
obtained. 
At the convergent point, the output of qcNN is of the form of kernel function 
of NTK. 
Because NTK is a nonlinear function of the above-mentioned covariance matrix 
composed of the quantum projection kernels, and because the computational cost 
of training is low, our qcNN can be regarded as a method to generate an effective 
quantum kernel with less computational complexity than the standard kernel method. 

\item
Because the NTK has an explicit form of covariance matrix, theoretical analysis 
on the training process and the convergent value of cost function is possible. 
As a result, based on this theoretical analysis on the cost function, 
we derive sufficient condition for our qcNN model to lower the cost function 
than some other full-classical models. 
Note that, when the size of quantum system is large, classical computers will 
have a difficulty to simulate the feature extraction process of the qcNN model; 
this may be a factor that leads to such superiority. 

\end{itemize}
In addition to the above theoretical investigations, we carry out thorough 
numerical simulations to evaluate the performance of the proposed qcNN 
model, as follows. 
\begin{itemize}

\item
The numerically computed time-evolution of cost function along the training 
process well agrees with the analytic form of time-evolution of cost (obtained 
under the assumption that NTK is constant and positive definite), for both the 
regression and classification problems, when the width of cNN is bigger than 100. 
This shows the validity of using NTK to analytically investigate the performance 
of the proposed qcNN. 

\item
The convergence speed becomes bigger (i.e., nearly the ideal exponentially-fast 
convergence is observed) and the value of final cost becomes smaller, when we 
make the width of cNN bigger. 
Moreover, we find that enough reduction of the training cost leads to the decrease of generalization 
error. 
That is, our qcNN has several desirable properties predicted by the NTK theory, 
which are indeed satisfied in many classical models. 

\item
Both the regression and classification performance largely depend on the choice 
of quantum circuit ansatz for data-encoding, which is reasonable in the sense 
that the proposed method is essentially a kernel method. 
Yet we found an interesting case where the ansatz with bigger-expressibility 
(due to containing some entangling gates) decreases the value of final cost 
lower than that achieved via the ansatz without entangling gates. 
This implies that the quantumness may have a power to enhance 
the performance of the proposed qcNN model, depending on the dataset or selected 
ansatz.

\item
The proposed qcNN model shows a clear advantage over full cNNs and qNNs for 
the problem of learning the quantum data-generating process. 
A particularly notable result is that, even with much less parameters (compared 
to the full cNNs) and smaller training cost (compared to the qNNs), the qcNN 
can execute the regression and the classification task with sufficient accuracy. 
Also, in terms of the generalization capability, the qcNN model shows much better 
performance than the others, mainly thanks to the inductive bias.

\end{itemize}

\subsection{Related works}

Before finishing this section, we address related works. 
Recently (after submitting a preprint version of this manuscript), the following 
studies on quantum NTK have been presented. 
Their NTK is defined for the cost function of the output state of a qNN. 
In Ref.~\cite{liu2022representation}, the authors studied 
the properties of the linear differential equation of the cost (which corresponds 
to Eq.~\eqref{equation:ft2} shown later), obtained under the assumption that the 
NTK does not change much in time. 
This idea was further investigated in the subsequent paper~\cite{PhysRevLett.130.150601}, 
showing in both theory and numerical simulation that the dynamics of cost 
exponentially decays when the number of parameters is large, i.e., when the 
system is within the over-parametrization regime, as suggested by the conventional 
classical NTK theory. 
This behaviour was also supported by numerical simulations provided in 
Ref.~\cite{shirai2021quantum}. 
Also, in Ref.~\cite{liu2022laziness}, a relation between their NTK and the 
vanishing gradient issue was discussed; 
that is, to satisfy the assumption that the NTK does not change in time, the qNN 
has to contain $O(4^n)$ parameters with $n$ the number of qubits, which actually 
has the same origin as the vanishing gradient issue. 
In Ref.~\cite{wang2022symmetric} the authors gave a method for mitigating this 
demanding requirement; they study the training dynamics in a space with 
effective dimension $d_{\rm eff}$ instead of the entire Hilbert space with 
dimension $2^n$, which as a result allows $O(d_{\rm eff}^2)$ parameters to 
guarantee the exponential convergence. 
All these studies focus on fully-quantum systems, while in this paper we focus 
on a class of classical-quantum hybrid systems where the tunable parameters 
are contained only in the classical part and the NTK is defined with respect 
to those parameters. 
A critical consequence due to this difference is that our NTK becomes 
time-invariant (Theorem 5) and the output function becomes Gaussian (Theorems 
3 and 4) in the over-parametrization regime, while these provable features were 
not reported in the above literature works. 
In particular, the time-invariancy is critical to guarantee the exponential 
convergence of the output function; as mentioned above, they rely on the 
assumption that the NTK does not change much in time. 
It may look like that our NTK is a fully classical object and as a result we 
are allowed to have such provable facts, but certainly it can extract features 
of the quantum part in the form of nonlinear function of the projected quantum 
kernel, as mentioned above.

\subsection{Structure of the paper}

The structure of this paper is as follows. 
Section II reviews the theory of NTK for cNNs. 
Section III begins with describing our proposed qcNN model, followed by showing 
some theorems. 
Also we discuss possible advantage of our qcNN over some other models. 
Section IV is devoted to give a series of numerical simulations. 
Section V concludes the paper.

\section{Preliminary: Classical neural tangent kernel theory}
\label{section:ntk-theory}

The NTK theory, which was originally proposed in \cite{NEURIPS2018_5a4be1fa}, 
offers a method for analyzing the dynamics of an infinitely-wide cNN under 
the gradient-descent-based training process. 
In particular, the NTK theory can be used for explaining why deep cNNs with 
much more parameters than the number of data (i.e., over-parametrized cNNs) 
work quite well in various machine learning tasks in terms of training error. 
We review the NTK theory in 
Sections from \ref{section:ntk-theory-problem-setting} to 
\ref{SECTION-ntk-consequence}. 
Importantly, the NTK theory can also be used to conjecture when cNNs may fail. 
As a motivation for introducing our model, we discuss one of the failure conditions 
of cNN in terms of NTK, in Section~\ref{section:ntk-theory-when-fail}.

\subsection{Problem settings of NTK theory}
\label{section:ntk-theory-problem-setting}

The NTK theory \cite{NEURIPS2018_5a4be1fa} focuses on supervised learning problems. 
That is, we are given $N_{D}$ training data ${(\bold{x}^a, y^a)}$ ($a=1,2,\cdots, N_D$), 
where $\boldx^a$ is an input vector and $y^a$ is the corresponding output; 
here we assume for simplicity that $y^a$ is a scalar, though the original NTK theory 
can handle the case of vector output. 
Suppose this dataset is generated from the following hidden (true) function $\fgoal$ 
as follows; 
\begin{equation}
    y^a = \fgoal(\boldx^a), \qquad \forall a.
\end{equation}
Then the goal is to train the model $\ft$, which corresponds to the output of a cNN, 
so that $\ft$ becomes close to $\fgoal$ in some measure, where $\thetat$ is the set 
of the trainable parameters at the iteration step $t$. 
An example of the measure that quantifies the distance between $\ft$ and $\fgoal$ is 
the mean squared error:
\begin{equation}
\label{equation:squared-loss}
    \mathcal{L}_t^{C} = \frac{1}{2}\sum_{a=1}^{N_D} (\ft\xa - \fgoal\xa )^2 = \frac{1}{2}\sum_{a=1}^{N_D} (\ft\xa - y^a )^2,
\end{equation}
which is mainly used for regression problems. 
Another example of the measure is the binary cross entropy:
\begin{equation}
\label{equation:binary-entropy}
    \mathcal{L}_t^{C} = -\sum_{a=1}^{N_D}\left(y^a \log \sigma_s (\ft\xa) + (1 - y_a) \log \sigma_s (\ft\xa)\right),
\end{equation}
which is mainly used for classification problems where $\sigma_s$ is the sigmoid 
function and $y^a$ is a binary label that takes either $0$ or $1$.

The function $\ft$ is constructed by a fully-connected network of $L$ layers. 
Let $n_\ell$ be the number of nodes (width) of the $\ell$-th layer (hence $\ell = 0$ 
and $\ell = L$ correspond to the input and output layers, respectively). 
Then the input $\bold{x}^a$ is converted to the output $\ft\xa$ in the following 
manner:
\begin{equation}
\begin{split}
    \boldalpha{0}{a} &= \bold{x}^a, \\
    \boldalpha{\ell}{a} &= \sigma(\tildealpha{\ell}{a}), \\
    \tildealpha{\ell + 1}{a}
       &= \frac{1}{\sqrt{n_{\ell}}}W^{(\ell)}\boldalpha{\ell}{a} + \xi b^{(\ell)}, \\
    \ft\xa &= \boldsymbol{\alpha}^{(L)}\xa,
\end{split}
\end{equation}
where $W^{(\ell)}\in \mathbf{R}^{n_{l} \times n_{l-1}}$ is the weighting matrix and 
$b^{(\ell)} \in \mathbf{R}^{n_{l}}$ is the bias vector in the $\ell$-th layer. 
Also $\sigma$ is the activation function that is differentiable. 
Note that the vector of trainable parameters $\thetat$ is now composed of all the 
elements of $\{W^{(\ell)}_{jk}\}$ and $b^{(\ell)}$. 
The parameters are updated by using the gradient descent algorithm
\begin{equation}
    \frac{\partial \theta_j (t)}{\partial t} = - \eta  \frac{\partial \mathcal{L}_t^{C}}{\partial \theta_j} = 
    -\eta\sum_a \frac{\partial \ft (\bold{x}^a)}{\partial \theta_j}
    \frac{\partial \mathcal{L}_t^{C}}{\partial \ft (\bold{x}^a)}, 
\end{equation}
where for simplicity we take the continuous-time regime in $t$. 
Also, $\eta$ is the learning rate and $\theta_j$ is the $j$-th parameter. 
All parameters, $\{W^{(\ell)}_{jk}\}$ and $b^{(\ell)}$, are initialized by 
sampling from the mutually independent normal Gaussian distribution.

\subsection{Definition of NTK}
\label{section:ntk-theory-definision}

NTK appears in the dynamics of the output function $\ft$, as follows. 
The time derivative of $\ft$ is given by
\begin{equation}
\label{equation:ft}
\begin{split}
    \frac{\partial \ft (\bold{x}))}{\partial t} &= \sum_{j}\frac{\partial \ft (\bold{x})}{\partial \theta_j}\frac{\partial \theta_j}{\partial t} \\
    &= -\eta \sum_{j,b}\frac{\partial \ft (\bold{x})}{\partial \theta_j}\frac{\partial \ft (\bold{x}^b)}{\partial \theta_j}\frac{\partial \mathcal{L}_t^{C}}{\partial \ft (\bold{x}^b)} \\
    &= -\eta \sum_b \kernelx{L}{\bold{x}}{\bold{x}^b} \frac{\partial \mathcal{L}_t^{C}}{\partial \ft (\bold{x}^b)}, 
\end{split}
\end{equation}
where $\kernel{L}$ is defined by 
\begin{equation}
\kernel{L} = \sum_{j}\frac{\partial \ft (\bold{x})}{\partial \theta_j}\frac{\partial \ft (\bold{x}^{\prime})}{\partial \theta_j}.
\end{equation}
The function $\kernel{L}$ is called the NTK. 
In the following, we will see that the trajectory of $\ft$ can be analytically 
calculated in terms of NTK in the infinite width limit $n_1,n_2,\cdots,n_{\ell-1} \rightarrow \infty$.

\subsection{Theorems}
\label{section:ntk-theory-theorems}

The key feature of NTK is that it converges to the time-invariant and 
positive-definite function $\ntk{L}$ in the infinite width limit, as shown below. 
Before stating the theorems on these surprising properties, let us show the 
following lemma about the distribution of $\fzero$:  

\begin{lemma}
\label{THEOREM-classical-cov}
{\bf (Proposition 1 in \cite{NEURIPS2018_5a4be1fa})}
With $\sigma$ as a Lipschitz nonlinear function, in the infinite width limit 
$n_{\ell} \rightarrow \infty$ for $1\leq\ell\leq L-1$, the output function 
at initialization, $\fzero$, obeys a centered Gaussian process whose covariance 
matrix $\cov{L}$ is given recursively by
\begin{equation}
\label{equation:classical-cov}
\begin{split}
     \cov{1} &= \frac{1}{n_0}\bold{x}^T \bold{x}^{\prime} + \xi^2,  %
    \\
  \cov{\ell + 1} &=\mathbf{E}_{h \sim \mathcal{N}\left(0, \bold{\Sigma}^{(\ell)}\right)}\left[\sigma(h(\bold{x})) \sigma\left(h\left(\bold{x}^{\prime}\right)\right)\right]+\xi^2,
\end{split}
\end{equation}
where the expectation is calculated by averaging over the centered Gaussian 
process with the covariance $\bold{\Sigma}^{(\ell)}$.
\end{lemma}

The proof can be found in Appendix A.1 of \cite{NEURIPS2018_5a4be1fa}. 
Note that the expectation for an arbitrary function 
$z(h(\bold{x}), h(\bold{x}^{\prime}))$ can be computed as
\begin{equation}
\mathbf{E}_{h \sim \mathcal{N}\left(0, \Sigma^{(\ell)}\right)}\left[z(h(\bold{x}), h(\bold{x}^{\prime}))\right]
=
    \frac{1}{2\pi\sqrt{|\Tilde{\bold{\Sigma}}^{(\ell)}|}}\int dh(\bold{x})dh(\bold{x}^{\prime}) \exp\left(-\frac{1}{2}\bold{h}^T\left(\Tilde{\bold{\Sigma}}^{(\ell)}\right)^{-1}\bold{h}\right)
    z(h(\bold{x}), h(\bold{x}^{\prime})),
\end{equation}
where $\Tilde{\bold{\Sigma}}^{(\ell)}$ is the $2 \times 2$ matrix
\begin{equation}
\Tilde{\bold{\Sigma}}^{(\ell)} = 
\left(
    \begin{array}{cc}
     \bold{\Sigma}^{(\ell)}(\bold{x}, \bold{x})   & \bold{\Sigma}^{(\ell)}(\bold{x}, \bold{x}^\prime) \\
     \bold{\Sigma}^{(\ell)}(\bold{x}^\prime, \bold{x})   & 
     \bold{\Sigma}^{(\ell)}(\bold{x}^\prime, \bold{x}^\prime)
    \end{array}
    \right),
\end{equation} the vector $\bold{h}$ is defined as $\bold{h}=\left(h(\bold{x}), h(\bold{x}^{\prime})\right)^T$,
and $|\Tilde{\bold{\Sigma}}^{(\ell)}|$ is the determinant of the matrix $\Tilde{\bold{\Sigma}}^{(\ell)}$.

From Lemma \ref{THEOREM-classical-cov}, the following theorem regarding NTK can be 
derived:

\begin{th.}
\label{THEOREM-classical-ntk}
{\bf (Theorem 1 in \cite{NEURIPS2018_5a4be1fa})}
With $\sigma$ as a Lipschitz nonlinear function, in the infinite width limit 
$n_{\ell} \rightarrow \infty$ for $1\leq\ell\leq L-1$, the neural tangent kernel 
$\kernel{L}$ converges to the time-invariant function $\ntk{L}$, which is given 
recursively by
\begin{equation}
    \begin{split}
            \ntk{1} &= \cov{1} =  \frac{1}{n_0}\bold{x}^T\bold{x}^\prime + \xi^2,  \\
    \ntk{\ell+1} &= \ntk{\ell}\covdot{\ell}+\cov{\ell+1},
    \end{split}
\end{equation}
where $\dot{\boldsymbol{\Sigma}}^{(\ell)}\left(\bold{x}, \bold{x}^{\prime}\right)=\mathbf{E}_{h \sim \mathcal{N}\left(0, \boldsymbol{\Sigma}^{(\ell)}\right)}\left[\dot{\sigma}(h(\bold{x})) \dot{\sigma}\left(h\left(\bold{x}^{\prime}\right)\right)\right]$ 
and $\dot{\sigma}$ is the derivative of $\sigma$. 
\end{th.}

Note that, by definition, the matrix $(\ntkx{L}{\bold{x}^a}{\bold{x}^b})$ is 
symmetric and positive semi-definite. 
In particular, when $L \geq 2$, the following theorem holds:

\begin{th.}
\label{THEOREM-positive-ntk}
{\bf (Proposition 2 in \cite{NEURIPS2018_5a4be1fa})}
With $\sigma$ as a Lipschitz nonlinear function, the kernel $\ntk{L}$ is positive 
definite when $L\geq 2$ and the input vector $\bold{x}$ is normalized as 
$\bold{x}^T \bold{x} = 1$.
\end{th.}

The above theorems on NTK in the infinite width limit can be utilized to analyze the 
trajectory of $\ft$ as shown in the next subsection.

\subsection{Consequence of Theorem \ref{THEOREM-classical-ntk} and Theorem \ref{THEOREM-positive-ntk}}
\label{SECTION-ntk-consequence}

From Theorems \ref{THEOREM-classical-ntk} and \ref{THEOREM-positive-ntk}, 
in the infinite width limit, the differential equation \eqref{equation:ft} can be 
exactly replaced by 
\begin{equation}
\label{equation:ft2}
    \frac{\partial \ft (\bold{x})}{\partial t} = -\eta \sum_b \ntkx{L}{\bold{x}}{\bold{x}^b} \frac{\partial \mathcal{L}_t^{C}}{\partial \ft (\bold{x}^b)}.
\end{equation}
The solution depends on the form of $\mathcal{L}_t^{C}$; of particular importance 
is the case when $\mathcal{L}_t^{C}$ is the mean squared loss. 
In our case \eqref{equation:squared-loss}, the functional derivative 
of the mean squared loss is given by 
\begin{equation}
\label{equation:functional-derivative in MSL case}
    \frac{\partial \mathcal{L}_t^{C}}{\partial \ft (\bold{x}^b)} 
        = \ft (\bold{x}^b) - y^b,
\end{equation}
and then we obtain the ordinary linear differential 
equation by substituting \eqref{equation:functional-derivative 
in MSL case} for \eqref{equation:ft2}.
This equation can be solved analytically~\cite{butcher2016numerical} at each 
data points as
\begin{equation}
\label{equation:theoretical-solution}
    \ft\xa=\sum_{j,b} V_{aj}^T\left(V_{jb}\fzero\xb - V_{jb}y^b\right) e^{-\eta \lambda_j t} 
             + y^a, 
\end{equation}
where $V=(V_{jb})$ is the orthogonal matrix that diagonalizes $\ntk{L}$ as
\begin{equation}
\label{equation:diagonalize}
     \sum_{a=1}^{N_D}\sum_{b=1}^{N_D}V_{ja}\ntkx{L}{\bold{x}^a}{\bold{x}^b}V_{bk}^{T} = \lambda_j \delta_{jk}.
\end{equation}
The eigenvalues $\lambda_j$ are non-negative, because $\ntk{L}$ is positive 
semi-definite.

When the conditions of Theorem \ref{THEOREM-positive-ntk} are satisfied, then 
$\ntk{L}$ is positive definite and accordingly $\lambda_j > 0$ holds for all $j$. 
Thus in the limit $t\rightarrow \infty$, the solution 
\eqref{equation:theoretical-solution} states that $\ft \xa = y_a$ holds for all $a$; 
namely, the value of the cost $\mathcal{L}_t^{C}$ reaches the global minimum 
$\mathcal{L}_t = 0$. 
This fine convergence to the global minimum explains why the over-parameterized 
cNN can be successfully trained.

We can also derive some useful theoretical formula for general $\bold{x}$. 
In the infinite width limit, from Eqs. \eqref{equation:ft2}, 
\eqref{equation:functional-derivative in MSL case}, and 
\eqref{equation:theoretical-solution} we have 
\begin{align}
    \frac{\partial \ft(\bold{x})}{\partial t} &= -\eta\sum_b \ntkx{L}{\bold{x}}{\bold{x}^b}(\ft(\bold{x}^b)-y^b) \\
    &= -\eta\sum_{b,c,j} \ntkx{L}{\bold{x}}{\bold{x}^b} V_{bj}^T(V_{jc}\fzero(\bold{x}^c)-V_{jc}y^c)e^{-\eta\lambda_j t}.
\end{align}
This immediately gives 
\begin{equation}
\label{equation:final-prediction-formula}
     \ft(\bold{x}) = -\sum_{b,c,j} \ntkx{L}{\bold{x}}{\bold{x}^b}V_{bj}^T D_j (V_{jc}\fzero(\bold{x}^c)-V_{jc}y^c),
\end{equation}
where 
\begin{equation}
    D_j = \left\{\begin{array}{clc}
         & (1-e^{-\eta\lambda_j t})/\lambda_{j} & (\lambda_j > 0)\\
         & \qquad\eta t &(\lambda_j=0)
    \end{array}
    \right..
\end{equation}
Now, if the initial parameters $\boldsymbol{\theta}(0)$ are randomly chosen 
from a centered Gaussian distribution, the average of $\ft(\bold{x})$ over such 
initial parameters is given by
\begin{align}
    \langle \ft(\bold{x}) \rangle 
        = \sum_{b,c,j}\ntkx{L}{\bold{x}}{\bold{x}^b}V_{bj}^T D_j V_{jc}y^c.
\end{align}

The formula \eqref{equation:final-prediction-formula} can be used for predicting 
the output for an unknown data, but it requires $O(N_D^3)$ computation to have 
$V$ via diagonalizing NTK, which may be costly when the number of data is large. 
To the contrary, in the case of cNN, the computational cost for its training is 
$O(N_D N_P)$, where $N_P$ is the number of parameters in cNN. 
Thus, if $N_D$ is so large that $O(N_D^3)$ classical computation is intractable, we can use the finite width cNN with 
$N_P \leq O(N_D)$, rather than \eqref{equation:final-prediction-formula} as 
a prediction function. 
In such case, the NTK theory can be used as theoretical tool for analyzing the 
behaviour of cNN.

Finally, let us consider the case where the cost is given by the binary cross 
entropy \eqref{equation:binary-entropy}; the functional derivative in this case 
is given by 
\begin{align}
    \frac{\partial \mathcal{L}_t^{C}}{\partial \ft(\bold{x}^a)} &= -y^a \frac{\dot{\sigma}_s(\ft(\bold{x}^a))}{\partial \ft(\bold{x}^a)} - (1 - y^a)\frac{-\dot{\sigma}_s(f(\bold{x}^a))}{1-\dot{\sigma}_s(\ft(\bold{x}^a))} \nonumber\\
    \label{equation:derivative}
    &= -y^a + \sigma(f(\bold{x}^a)),
\end{align}
where in the last line we use the derivative formula for the sigmoid function:
\begin{equation}
    \dot{\sigma}_s(q) = \left(1-\sigma_s(q)\right)\sigma_s(q).
\end{equation}
By substituting \eqref{equation:derivative} into \eqref{equation:ft2}, we obtain
\begin{equation}
    \ft\xa = -\eta \int_0^{t}dt^{\prime}\sum_b \ntkx{L}{\bold{x}^a}{\bold{x}^b}\left(-y^b + \sigma(f_{\boldsymbol{\theta}(t^\prime)}\xa)\right),
\end{equation}
and similarly for general input $\bold{x}$
\begin{equation}
    \ft(\bold{x}) = -\eta \int_0^{t}dt^{\prime}\sum_b \ntkx{L}{\bold{x}}{\bold{x}^b}\left(-y^b + \sigma(f_{\boldsymbol{\theta}(t^\prime)}\xa)\right).
\end{equation}
These are not linear differential equations and thus cannot be solved analytically, unlike the mean squared error case; but we can numerically solve them by 
using standard ordinary differential equation tools \cite{butcher2016numerical}.

\subsection{When may cNN fail?}
\label{section:ntk-theory-when-fail}

The NTK theory tells that, as long as the condition of Theorem \ref{THEOREM-positive-ntk} holds, the cost function converges to the global minimum in the limit $t\rightarrow \infty$. 
However in practice we must stop the training process of cNN at a finite time 
$t=\tau$. 
Thus, the speed of convergence is also an important factor for analyzing the behaviour 
of cNN. 
In this subsection we discuss when cNN may fail in terms of the convergence speed. 
We discuss the case when the cost is the mean squared loss.

Recall now that the speed of convergence depends on the eigenvalues 
$\{\lambda_j\}_{j=1}^{N_D}$. 
If the minimum of the eigenvalues, $\lambda_{\rm min}$, is sufficiently larger than 0, 
the cost function quickly converges to the global minimum in the number of iteration 
$O(1/\lambda_{\rm min})$. 
Otherwise, the speed of convergence is not determined only by the spectrum of the 
eigenvalues, but the other factors in \eqref{equation:theoretical-solution} need to 
be taken into account; 
actually many of the reasonable settings correspond to this case \cite{arora2019fine}, 
and thus we will consider this setting in the following.

First, the formula \eqref{equation:theoretical-solution} can be rewritten as
\begin{equation}
\label{cNN failure section w_j}
    w_j(t)=\left(w_{j}(0)-g_{j}\right) e^{-\eta \lambda_j t}+g_{j}, 
\end{equation}
where $w_j(t) = \sum_a V_{ja}\ft(x_a)$ and $g_j = \sum_a V_{ja}y_a$. 
Let us assume that we stop the training at $t=\tau < O(1/\lambda_{\rm min})$. 
With $S_{\eta \tau}=\{j|\lambda_j < 1/\eta\tau, 1\leq j \leq N_D\}$, 
if we approximate the exponential function as 
\begin{equation}
\label{eq:exp-approx}
e^{-\eta\lambda_j \tau} \simeq
\left\{ 
    \begin{array}{cc}
              1 & {\rm if}\ j\in S_{\eta \tau} \\
         0  & {\rm otherwise}
    \end{array}
\right.
,
\end{equation}
then we obtain
\begin{equation}
\label{equation:solution}
w_j(\tau) \simeq \left\{
\begin{array}{cl}
    w_j(0) & {\rm if}\ j\in S_{\eta \tau}\\
    g_j & {\rm otherwise}
\end{array}
\right.
.
\end{equation}

By using the same approximation, the cost function at the iteration step $\tau$ 
can be calculated as
\begin{equation}
\begin{split}
\mathcal{L}_\tau^C \equiv \frac{1}{N_D}\sum_{a=1}^{N_D}(\ftau\xa - y^a)^2& =   \frac{1}{N_D}\sum_{a=1}^{N_D}\left[\sum_{j=1}^{N_D}V^{T}_{aj}\left( w_j(\tau) - g_j\right) \right]^2 \\
&\simeq \frac{1}{N_D}\sum_{a=1}^{N_D}\left(\sum_{j\in S_{\eta\tau}}V^{T}_{aj}(w_j(0) - g_j)\right)^2 \\
&= \frac{1}{N_D}\sum_{j\in S_{\eta\tau}} w_j(0)^2 + \frac{1}{N_D}\sum_{j\in S_{\eta\tau}}g_j^2 - \frac{2}{N_D}\sum_{j\in S_{\eta\tau}}w_j(0)g_j.
\end{split}
\end{equation}
Since $w_j(0)$ is the sum of centered Gaussian distributed variables, $w_j(0)$ 
also obeys the centered Gaussian distribution with covariance:
\begin{equation}
    \begin{split}
    \langle w_j(0)w_k(0) \rangle &= \sum_{a,b}V_{ja}V_{kb} \langle \fzero\xa\fzero\xb \rangle \\
    &= \sum_{a,b}V_{ja} \covx{L}{\bold{x}^a}{\bold{x}^b}V_{bk}^{T}.
    \end{split}
\end{equation}
Thus, we have 
\begin{equation}
\label{equation:average-cost}
    \langle \mathcal{L}_\tau^C\rangle \simeq \frac{1}{N_D}\sum_{j\in S_{\eta\tau}}\sum_{b,c}V_{jb} \covx{L}{\bold{x}^b}{\bold{x}^c}V_{cj}^{T} +  \frac{1}{N_D}\sum_{j\in S_{\eta\tau}}g_j^2.
\end{equation}
Since the covariance matrix can be diagonalized with an orthogonal matrix 
$V^{\prime}$ as
\begin{equation}
    V^{\prime}_{jb}\covx{L}{\bold{x}^b}{\bold{x}^c}V^{\prime T}_{ck} 
       = \lambda_j^{\prime} \delta_{jk},
\end{equation}
the first term of Eq.~\eqref{equation:average-cost} can be rewritten as
\begin{equation}
\label{equation:first-term-cost}
     \frac{1}{N_D}\sum_{j\in S_{\eta\tau}}\sum_{b,c}V_{jb} \covx{L}{\bold{x}^b}{\bold{x}^c}V_{cj}^{T}
     = \frac{1}{N_D}\sum_{j\in S_{\eta\tau}}\sum_{k=1}^{N_D}\lambda_k^{\prime}(\bold{v}^{\prime}_k \cdot \bold{v}_j)^2,
\end{equation}
where $\bold{v}_j = \{V_{ja}\}_{a=1}^{N_D}$ and $\bold{v}_j^{\prime} = \{V^{\prime}_{ja}\}_{a=1}^{N_D}$. 
Also, the second term of \eqref{equation:average-cost} can be written as
\begin{equation}
    \frac{1}{N_D}\sum_{j\in S_{\eta\tau}}g_j^2 = \frac{1}{N_D}\sum_{j\in S_{\eta\tau}}(\bold{y} \cdot \bold{v}_j)^2,
\end{equation}
where $\bold{y}$ is the label vector defined by $\bold{y}=\{y^a\}_{a=1}^{N_D}$. 
Thus, we have 
\begin{equation}
\label{equation:classical-problem}
    \langle \mathcal{L}_{\tau}^{C}\rangle \simeq \frac{1}{N_D}\sum_{j\in S_{\eta\tau}}\sum_{k=1}^{N_D}\lambda_k^{\prime}(\bold{v}^{\prime}_k \cdot \bold{v}_j)^2 +  \frac{1}{N_D}\sum_{j\in S_{\eta\tau}}(\bold{y} \cdot \bold{v}_j)^2.
\end{equation}
The cost $\mathcal{L}_\tau^{C}$ becomes large, depending on the values of the 
first and the second terms, characterized as follows: 
(i) the first term becomes large if the 
eigenvectors of $\covx{L}{\bold{x}^b}{\bold{x}^c}$  with respect to large eigenvalues align with the eigenvectors of $\ntkx{L}{\bold{x}^b}{\bold{x}^c}$ with respect to small eigenvalues and (ii) the 
second term becomes large if the label vector aligns with the eigenvectors of $\ntkx{L}{\bold{x}^b}{\bold{x}^c}$ with respect to small eigenvalues. 
Of particular importance is the condition where the latter statement (ii) applies. 
Namely, the cNN cannot be well optimized in a reasonable time if we use a dataset 
whose label vector aligns with the eigenvectors of $\ntkx{L}{\bold{x}^b}{\bold{x}^c}$ 
with respect to small eigenvalues. 
If such a dataset is given to us, therefore, an alternative method that may 
outperform the cNN is highly demanded, which is the motivation of introducing 
our model.

\vspace{0.4cm}
\noindent{\bf Remark 1:} 
If some noise is added to the label of the training data, we need not aim to decrease 
the cost function toward precisely zero. 
For example, when the noise vector $\boldsymbol{\epsilon}$ is appended to the true 
label vector $\Tilde{\bold{y}}$ in the form $\bold{y}=\Tilde{\bold{y}}+\boldsymbol{\epsilon}$, it may be favorable to stop the 
optimization process at time $t=\tau$ before 
$\sum_{j\in S_{\eta\tau}}(\boldsymbol{\epsilon}\cdot\bold{v})^2$ becomes small, 
for avoiding the overfitting to the noise; actually in the original NTK paper 
\cite{NEURIPS2018_5a4be1fa} the idea of avoiding the overfitting by using early 
stopping is mentioned. 
In this case, instead of $\sum_{j\in S_{\eta\tau}}(\bold{y}\cdot\bold{v})^2$, 
we should aim to decrease the value of 
$\sum_{j\in S_{\eta\tau}}(\Tilde{\bold{y}}\cdot\bold{v})^2$, to construct a prediction function that has a good generalization ability.

\section{Proposed model}

\begin{figure}[h]
  \centering
  \includegraphics[width=400pt]{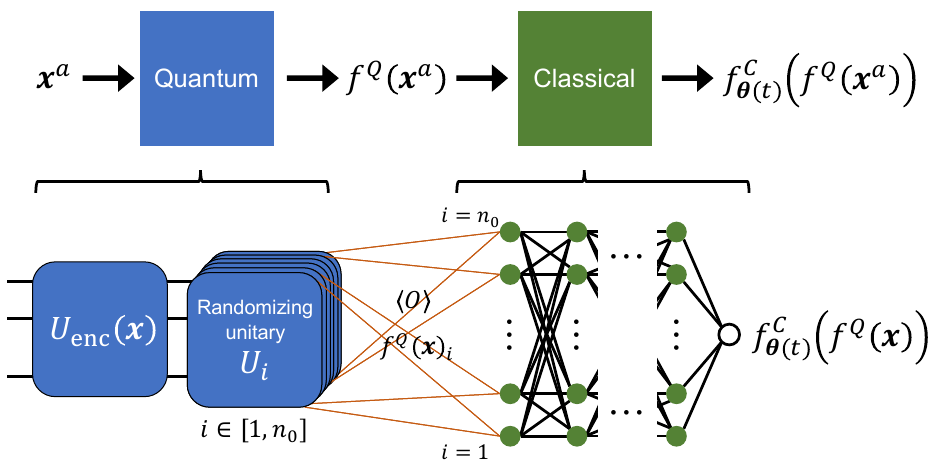}
  \caption{
  The overview of the proposed qcNN model. 
  The first quantum part is composed of the encoding unitary 
  $U_{\rm enc}(\boldx^a)$ for the data $\boldx^a$ followed by the random 
  unitary $U_i$ and measurement of an observable $O$ for extracting a feature 
  of the quantum state, $\fq \xa_i$. 
  We run $n_0$ different quantum circuits to construct a feature vector 
  $\boldfq=\fqarray$, which is the input vector to the classical part 
  composed of $n_0$-nodes multi-layered NN.} 
  \label{FIG_model}
\end{figure}

In this section, we introduce our qcNN model for supervised learning, which is 
theoretically analyzable using the NTK theory. 
Before describing the detail, we summarize the notable point of this qcNN. 
This qcNN is a concatenation of a quantum circuit followed by a cNN, as 
illustrated in Fig.~\ref{FIG_model}. 
Likewise the classical case shown in Section~\ref{SECTION-ntk-consequence}, 
we obtain the time-invariant NTK in the infinite width limit of the cNN part, 
which allows us to theoretically analyze the behaviour of the entire system. 
Importantly, NTK in our model coincides with a certain quantum kernel 
computed in the quantum data-encoding part. 
This means that the output of our qcNN can 
represent functions of quantum states defined on the quantum feature space 
(Hilbert space); hence, if the quantum encoder is designed appropriately, 
our model may have advantage over purely classical systems. 
In the following, we discuss the detail of our model from  
Section~\ref{section:detail-proposed-model} to 
Section~\ref{section:feature-proposed-model}, and discuss possible advantage 
in Section~\ref{section:qntk-advantage}.

\subsection{qcNN model}
\label{section:detail-proposed-model}

We consider the same supervised learning problem discussed in Section II. 
That is, we are given $N_{D}$ training data ${(\bold{x}^a, y^a)}$ 
($a=1,2,\cdots, N_D$) generated from the hidden function $\fgoal$ satisfying 
\begin{equation}
    y^a = \fgoal(\boldx^a), \qquad \forall a.
\end{equation}
Then the goal is to train the model function $\ft$ so that $\ft$ becomes closer 
to $\fgoal$ in some measure, by updating the vector of parameters $\thetat$ as 
a function of time $t$. 
Our qcNN model $\ft$ is composed of the quantum part ${\bold f}^Q$ and the classical 
part $\fc$, which are concatenated as follows: 
\begin{equation}
    \ft = \fc \circ {\bold f}^Q.
\end{equation}
Only the classical part has trainable parameters in our model as will be stated 
later, and thus the subscript $\thetat$ is placed only on the classical part.

The quantum part first operates the $n$-qubits quantum circuit (unitary operator) 
$\uenc$ that loads the classical input data $\boldx^a$ into the quantum state 
in the manner $|\psi(\boldx^a)\rangle = U_{\rm enc}(\boldx^a)|0\rangle^{\otimes n}$. 
We then operate a random unitary operator $U_i$ on the quantum state 
$|\psi(\bold{x}^a)\rangle$ and finally measure an observable $O$ to have 
the expectation value 
\begin{equation}
    \fq \xa_i 
    = \langle\psi(\xa)|U_i^{\dagger} O U_i  |\psi(\xa)\rangle
    = \langle 0|^{\otimes n}\uencxa^{\dagger} U_i^{\dagger} O U_i \uencxa|0\rangle^{\otimes n}. 
\end{equation}
We repeat this procedure for $i=1,\ldots,n_0$ and collect these quantities to 
construct the $n_0$-dimensional vector $\boldfq=\fqarray$, which is the output 
of the quantum part of our model. 
The randomizing process corresponds to extracting features of 
$|\psi(\bold{x}^a)\rangle$, likewise the machine learning method using the 
classical shadow tomography \cite{huang2020predicting,huang2022provably}; 
but our method does not construct a tomographic density matrix (called the 
snapshot) but directly construct the feature vector $\boldfq$ which will be 
further processed in the classical part. 
Note that, as shown later, we will make $n_0$ bigger sufficiently so that the 
NTK becomes time-invariant and thereby the entire dynamics is analytically 
solvable. 
Hence it may look like that the procedure for constructing the $n_0$-dimensional 
vector $\boldfq$ is inefficient, but practically a modest number of $n_0$ is 
acceptable, as demonstrated in the numerical simulation in Section~\ref{section:exp_qdata}.

In this paper, we take the following setting for each component. 
The classical input data $\boldx^a$ is loaded into the $n$-qubits quantum 
state through the encoder circuit $\uenc$. 
Ideally, we should design the encoder circuit $\uenc$ so that it reflects the 
hidden structure (e.g., symmetry) of the training data, as suggested in \cite{NEURIPS2021_69adc1e1,ragone2022representation}; the numerical simulation in Section~\ref{section:exp_qdata} considers this case. 
As for the randomizing unitary operator $U_i$, it is of the tensor product form: 
\begin{equation}
    U_i = U_i^1 \otimes U_i^2 \otimes \cdots U_i^{n_Q}, 
\end{equation}
where $m$ is an integer called the {\it locality}, and we assume that $n_Q=n/m$ 
is an integer. 
Each $U_i^{k}$ $(k=1,2,\cdots,n_Q)$ is independently sampled from unitary 2-designs and is fixed during the training. 
Note that a unitary 2-design is implementable on a circuit with 
the number of gates $O(m^2)$ \cite{dankert2009exact}. 
Lastly, the observable $O$ is the sum of $n_Q$ local operators:
\begin{equation}
    O = \sum_{k=1}^{n_Q}I_{(k-1)m}\otimes \mathcal{O} \otimes I_{(n_Q-k)m}, 
\end{equation}
where $I_{u}$ is the $2^u$-dimensional identity operator and $\mathcal{O}$ 
is a $2^m$-dimensional traceless operator.

Next we describe the classical part, $\fc$. 
This is a cNN that takes the vector $\boldfq$ as the input and returns the output  
$\fc(\bold{f}_Q)$; therefore, $\ft\xa = \fc(\boldfq)$. 
We implement $\fc$ as an $L$-layers fully connected cNN, which is the same as 
that introduced in Section~\ref{section:ntk-theory}:
\begin{equation}
\label{equation:quantum-network-propagation}
\begin{split}
    \boldalpha{0}{a} &= \boldfq, \\
    \boldalpha{\ell}{a} &= \sigma(\tildealpha{\ell}{a}), \\
    \tildealpha{\ell + 1}{a}&= \frac{1}{\sqrt{n_{\ell}}}W^{(\ell)}\boldalpha{\ell}{a} + \xi b^{(\ell)}, \\
    \fc(\bold{f}\xa) &= \boldsymbol{\alpha}^{(L)}\xa,  
\end{split}
\end{equation}
where $\ell=0,1,\cdots,L-1$. 
As in the case of cNN studied in Section~\ref{section:ntk-theory}, $W^{(\ell)}$ 
is the $n_{\ell +1}\times n_{\ell}$ weighting matrix and $b^{(\ell)}$ is the 
$n_{\ell}$-dimensional bias vector; 
each element of $W$ and $b^{(\ell)}$ are initialized by sampling from the 
mutually independent normal Gaussian distributions.

The parameter $\thetat$ is updated by the gradient descent algorithm
\begin{equation}
\label{equation:gradient-descent}
    \frac{\partial\theta_p(t)}{\partial t} = -\eta \frac{\partial \mathcal{L}^Q_{t}}{\partial \theta_p(t)}, 
\end{equation}
where $\mathcal{L}^Q_{t}$ is the cost function that reflects a distance 
between $\ft$ and $\fgoal$. 
Also $\eta$ is the learning rate and $\theta_p(t)$ $(p=1,2,\cdots,P)$ is the 
$p$-th element of $\thetat$ that corresponds to the elements of 
$W^{(1)}, W^{(2)}, \cdots, W^{(L-1)}$ and $b^{(1)}, b^{(2)}, \cdots, b^{(L-1)}$. 
The task of updating the parameters only appears in the classical part, which 
can thus be performed by applying some established machine learning solver 
given the $N_D$ training data $\{(\boldx^a, y^a) \}$ $(a=1,2,\cdots,N_D)$, 
cNN $\fc$, and the cached output from the quantum part at initialization.

\subsection{Quantum neural tangent kernel}

As proven in Section~\ref{section:ntk-theory}, when the parameters are updated 
via the gradient descent method 
\eqref{equation:gradient-descent}, the output function $\ft$ changes in time 
according to 
\begin{equation}
\label{equation:theoretical-formula-q}
\frac{\partial \ft(\bold{x})}{\partial t}= -\eta \sum_{a=1}^{N_D}K_Q\left(\bold{x}, \bold{x}^a, t\right) \frac{\partial \mathcal{L}^Q_{t}}{\partial \ft\left(\bold{x}^a\right)}. 
\end{equation}
Here $K_Q(\bold{x}, \bold{x}^{\prime}, t)$ is the {\it quantum neural tangent 
kernel (QNTK)}, defined by
\begin{equation}
\label{formula:quantum-enhanced-ntk}
    K_Q(\bold{x}, \bold{x}^{\prime}, t) = \sum_{p=1}^{P}\frac{\partial \ft(\bold{x})}{\partial \theta_p(t)}\frac{\partial \ft(\bold{x}^{\prime})}{\partial \theta_p(t)}.
\end{equation}
It is straightforward to show that $K_Q(\bold{x}, \bold{x}^{\prime}, t)$ is 
positive semi-definite. 
We will see the reason why we call $K_Q(\bold{x}, \bold{x}^{\prime}, t)$ as 
the {\it quantum} neural tangent kernel in the next subsection.

\subsection{Theorems}
\label{section:feature-proposed-model}

We begin with the theorem stating the probability distribution of the output 
function $\fzero$ in the case $L=1$; this setting shows how a quantum kernel 
appears in our model, as follows.

\begin{th.}
\label{theorem:covqone}
With $\sigma$ as a Lipschitz function, for $L=1$ and in the limit $n_0\xrightarrow{}\infty$, the output function $\fzero$ is a centered Gaussian 
process whose covariance matrix $\covq{1}$ is given by 
\begin{equation}
    \covq{1} = \frac{{\rm Tr}(\mathcal{O}^2)}{2^{2m}-1}\sum_{k=1}^{n_Q}\left({\rm Tr}(\rhok{x}\rhok{x^{\prime}}) - \frac{1}{2^m}\right) + \xi^2.
\end{equation}
Here $\rho_x^k$ is the reduced density matrix defined by
\begin{equation}
\label{equation:projected-density-matrix-1}
    \rhok{x} = {\rm Tr}_k\left(\uenc(\boldx)|0\rangle^{\otimes n}\langle 0|^{\otimes n}\uenc(\boldx)^{\dagger}\right), 
\end{equation}
where ${\rm Tr}_k$ is the partial trace over the entire Hilbert space except 
from the $(km-m)$-th qubit to the $(km-1)$-th qubit.
\end{th.}

The proof is found in Appendix \ref{section:theorem-three}. 
Note that the term $\sum_{k=1}^{n_Q} {\rm Tr}(\rhok{x}\rhok{x^{\prime}})$ 
coincides with one of the projected quantum kernels introduced in
\cite{huang2021power} with the following motivation. 
That is, when the number of qubits (hence the dimension of Hilbert space) 
becomes large, the Gram matrix composed of the inner product between pure 
states, ${\rm Tr}(\rho_\bold{x} \rho_{\bold{x}^{\prime}}) 
= |\langle \psi(\bold{x})|\psi(\bold{x}^{\prime})\rangle|^2$, becomes close 
to the identity matrix under certain type of feature map~\cite{huang2021power, thanasilp2022exponential, suzuki2022quantum}, meaning that there is no quantum advantage in using 
this kernel. 
The projected quantum kernel may cast as a solution for this problem; that is, 
by projecting the density matrix in a high-dimensional Hilbert space to a 
low-dimensional one as in \eqref{equation:projected-density-matrix-1}, the 
Gram matrix of kernels defined by the inner product of projected density 
matrices can take some quantum-intrinsic structure which largely differs from 
the identity matrix.

The covariance matrix $\covq{1}$ inherits the projected quantum kernel, which 
can be more clearly seen from the following corollary:

\begin{co.}
The covariance matrix obtained in the setting of Theorem \ref{theorem:covqone} is of the form 
\begin{equation}
        \covq{1} 
          = \frac{{\rm Tr}(\mathcal{O}^2)}{2^{2m}-1}\sum_{k=1}^{n_Q}{\rm Tr}(\rhok{x}\rhok{x^{\prime}}), 
\end{equation}
if $\xi$ is set to be
\begin{equation}
    \xi = \sqrt{\frac{n_Q{\rm Tr}(\mathcal{O}^2)}{(2^{2m}-1)2^m}}.
\end{equation}
\end{co.}

Namely, $\covq{1}$ is exactly the projected quantum kernel up to the constant factor, if we suitably 
choose the coefficient of the bias vector given in Eq.~\eqref{equation:quantum-network-propagation}.

Based on result in the case of $L=1$, we can derive the following 
Theorem~\ref{theorem:quantum-gaussian-process} and Theorem~\ref{theorem:ntkq}.
First, the distribution of $\fzero$ when $L > 1$ can be recursively computed 
as follows.

\begin{th.}
\label{theorem:quantum-gaussian-process}
With $\sigma$ as a Lipschitz function, for $L >1$ and in the limit 
$n_0, n_1, \cdots, n_{L-1}\xrightarrow{} \infty$, $\fzero$ is a centered Gaussian process whose 
covariance matrix $\covq{L}$ is given recursively by 
\begin{equation}
\label{equation:quantum-cov-1}
\begin{split}
  \covq{1} &= \frac{{\rm Tr}(\mathcal{O}^2)}{2^{2m}-1}
          \sum_{k=1}^{n_Q}\left({\rm Tr}(\rhok{x}\rhok{x^{\prime}}) - \frac{1}{2^m}\right) + \xi^2, \\
  \covq{\ell + 1} &=\mathbf{E}_{h \sim \mathcal{N}\left(0, \Sigma_Q^{(\ell)}\right)}\left[\sigma(h(\bold{x})) \sigma\left(h\left(\bold{x}^{\prime}\right)\right)\right]+\xi^2, 
\end{split}
\end{equation}
where the expectation value is calculated by averaging over the centered Gaussian process with 
covariance matrix $\Sigma_Q^{(\ell)}$. 
\end{th.}  
\noindent The proof is found in Appendix~\ref{section:proof-theorem-four}. 
Note that the only difference between the quantum case \eqref{equation:quantum-cov-1} and the 
classical case \eqref{equation:classical-cov} is that the covariance matrix corresponding to the first 
layer in the entire network.

The infinite width limit of the QNTK can be also derived in a similar manner as Theorem \ref{THEOREM-classical-ntk}, as follows.

\begin{th.}
\label{theorem:ntkq}
With $\sigma$ as a Lipschitz function, in the limit $n_0, n_1, \cdots, n_{L-1}\xrightarrow{} \infty$, 
the QNTK $K_Q(\bold{x}, \bold{x}^{\prime}, t)$ converges to the time-invariant function $\ntkq{L}$, 
which is given recursively by
\begin{align}
\label{equation:qntk-infinite-1}
\begin{split}
    \ntkq{1} &= \covq{1} =  \frac{{\rm Tr}(\mathcal{O}^2)}{2^{2m}-1}\sum_{k=1}^{n_Q}\left({\rm Tr}(\rhok{\bold{x}}\rhok{\bold{x}^{\prime}}) - \frac{1}{2^m}\right) + \xi^2,  \\
    \ntkq{\ell+1} &= \ntkq{\ell}\dot{\boldsymbol{\Sigma}}_Q^{(\ell)}\left(\bold{x}, \bold{x}^{\prime}\right)+\boldsymbol{\Sigma}_Q^{(\ell+1)}\left(\bold{x}, \bold{x}^{\prime}\right), 
\end{split}
\end{align}
where $\dot{\boldsymbol{\Sigma}}_Q^{(\ell)}\left(\bold{x}, \bold{x}^{\prime}\right)=\mathbf{E}_{h \sim \mathcal{N}\left(0, \boldsymbol{\Sigma}_Q^{(\ell)}\right)}\left[\dot{\sigma}(h(\bold{x})) \dot{\sigma}\left(h\left(\bold{x}^{\prime}\right)\right)\right]$ and $\dot{\sigma}$ is the derivative of $\sigma$.
\end{th.}

\noindent 
The proof is in Appendix~\ref{section:proof-five}. 
Note that the above two theorems can be proven with almost the same manner as 
in \cite{NEURIPS2018_5a4be1fa}.

When $L=1$, the QNTK directly inherits the structure of the quantum kernel, and 
this is the reason why we call $K_Q(\bold{x}, \bold{x}^{\prime}, t)$ the 
{\it quantum} NTK. 
Also, such inherited structure in the first layer propagates to the subsequent 
layers when $L>1$; 
the resulting kernel is then of the form of a nonlinear function of the 
projected quantum kernel. 
Considering the fact that designing an effective quantum kernel is in general 
quite nontrivial, it is useful for us to have a method to automatically generate 
a nonlinear kernel function appearing when $L>1$. 
Note that, when the ReLU activation function is used, the analytic form of 
$\ntkq{L}$ is recursively computable as shown in Appendix~\ref{section:relu-analytic}.

As in the classical case, Theorem~\ref{theorem:ntkq} is the key property that 
enables us to analytically study the training process of the qcNN. 
In particular, let us recall Theorem~2 and the discussion below Eq.~(15), 
showing the importance of positive semi-definiteness or definiteness of 
the kernel $\ntkq{L}$. 
(The positive semi-definiteness is trivial since $K_Q(x, x^{\prime}, t)$ is 
positive semi-definite.) 
Actually, we now have an analogous result to Theorem 2 as follows. 
\begin{th.}
\label{theorem:positive-ntkq}
For a non-constant Lipschitz function $\sigma$, QNTK $\ntkq{L}$ is positive definite unless there exists $\{c_a\}_{a=1}^{N_D}$ such that (i) $\sum_a c_a\rho_{\bold{x}^a}^k=\bold{0}$\ $(\forall k)$, $\sum_a c_a = 0$, and $c_a \neq 0\ (\exists a)$ or (ii) $\xi=0$, $\sum_a c_a\rho_{\bold{x}^a}^k=I_{m}/2^m$\ $(\forall k)$ and $\sum_a c_a = 1$.
\end{th.}
\noindent We give the proof in Appendix~\ref{section:qntk-positive}. 
Note that condition (i) can be interpreted as the data embedded 
reduced density matrices being linearly dependent, which can be avoided by 
removing redundant data. 
It is difficult to give a proper interpretation on the condition (ii), but 
it is still avoidable by setting $\xi$ larger than zero.

Based on the above theorems, we can theoretically analyze the learning process and moreover 
the resulting performance. 
In the infinite-width limit of cNN part, the dynamics of the output function $\ft (\bold{x})$ given by 
Eq.~\eqref{equation:theoretical-formula-q} takes the form 
\begin{equation}
\label{equation:ft2q}
    \frac{\partial \ft (\bold{x})}{\partial t} 
      = -\eta \sum_b \ntkqx{L}{\bold{x}}{\bold{x}^b} 
                      \frac{\partial \mathcal{L}_t^{Q}}{\partial \ft (\bold{x}^b)}. 
\end{equation}
Because the only difference between this dynamical equation and that for the 
classical case, Eq.~\eqref{equation:ft2}, is in the form of NTK, the discussion 
in Section~\ref{SECTION-ntk-consequence} can be directly applied. 
In particular, if the cost $\mathcal{L}_t^{Q}$ is the mean squared error \eqref{equation:squared-loss}, 
the solution of Eq.~\eqref{equation:ft2q} is given by 
\begin{equation}
\label{equation:theoretical-solution-q}
    \ft\xa=\sum_j V_{aj}^{QT}\left(V_{jb}^Q\fzero\xb - V_{jb}^Q y^b\right) e^{-\eta \lambda_j t} + y^a,
\end{equation}
where $V^Q$ is the orthogonal matrix that diagonalizes $\ntkq{L}$ as
\begin{equation}
\label{equation:diagonalize-q}
     \sum_{a=1}^{N_D}\sum_{b=1}^{N_D}V_{ja}^Q\ntkx{L}{\bold{x}^a}{\bold{x}^b}V_{bk}^{QT} 
        = \lambda_j^Q \delta_{jk}.
\end{equation}
$\{\lambda_j^Q\}$ are the eigenvalues of $\ntkq{L}$, which is generally 
positive semi-definite. 
If Theorem~6 holds, then $\ntkq{L}$ is positive definite or equivalently 
$\{\lambda_j^Q\}$ are all positive; then Eq.~\eqref{equation:theoretical-solution-q}
shows $\ft\xa \to y^a$ as $t\to\infty$ and thus the learning process perfectly 
completes. 
Note that, if the cost is the binary cross-entropy \eqref{equation:binary-entropy}, 
then we have 
\begin{equation}
\label{equation:binary}
     \ft\xa = -\eta \int_0^{t}dt^{\prime}\sum_b \ntkx{L}{\bold{x}^a}{\bold{x}^b}\left(-y^b + \sigma(f_{\boldsymbol{\theta}(t^\prime)}\xa)\right).
\end{equation}

\subsection{Possible advantage of the proposed model}
\label{section:qntk-advantage}

In this subsection, we discuss two scenarios where the proposed qcNN has possible 
advantage over other models.

\subsubsection*{Possible advantage over pure classical models}

First, we discuss a possible advantage of our qcNN over classical models. 
For this purpose, recall that our QNTK contains features of quantum states in the 
form of a nonlinear function of the projected quantum kernel, as proven in Theorem~5. 
Hence, under the assumption of the classical intractability for the projected quantum kernel~\cite{huang2021power}, our QNTK may also be a classically intractable object. 
As a result, the output function \eqref{equation:theoretical-solution-q} or 
\eqref{equation:binary} may potentially achieve the training error or the 
generalization error smaller than that any classical means cannot reach. 
Now, considering the fact that designing an effective quantum kernel is in general 
quite nontrivial, it is useful for us to have a NN-based method for synthesizing 
a nonlinear kernel function that really outperforms any classical means 
for a given task.

To elaborate on the above point, let us study the situation where a quantum 
advantage would appear in the training error. 
More specifically, we investigate the condition where
\begin{equation}
\label{equation:general-advantage-condition}
   \min_{\sigma\in F,L}\langle\mathcal{L}_\tau^C \rangle  > 
      \min_{\sigma\in F,L,U_{enc}}\langle\mathcal{L}_\tau^Q \rangle,
\end{equation}
holds.  Here we assume that the time $\tau$ is sufficiently large such that further training does not change the cost. Also, $F$ is the set of differentiable Lipschitz functions, $L$ is the number of layer of cNN, and the average is 
taken over the initial parameters. If \eqref{equation:general-advantage-condition} holds, we can say that our qcNN model is better than the pure classical model regarding the training error. 
To interpret the condition \eqref{equation:general-advantage-condition} analytically, 
let us further assume that the cost is the mean squared error.  
Then, the condition 
\eqref{equation:general-advantage-condition} is approximately rewritten by 
using Eq.~\eqref{equation:classical-problem} as
\begin{equation}
\label{equation:advantage-mse}
    \min_{\sigma\in F,L}\left\{ \sum_{j\in S_{\eta\tau}^C}\sum_{k=1}^{N_D}\lambda_k^{C\prime}(\bold{v}^{C\prime}_k \cdot \bold{v}_j^C)^2 + \sum_{j\in S_{\eta\tau}^C}(\bold{y} \cdot \bold{v}_j^C)^2\right\} > 
    \min_{\sigma\in F,L,U_{enc}} \left\{\sum_{j\in S_{\eta\tau}^Q}\sum_{k=1}^{N_D}\lambda_k^{Q\prime}(\bold{v}^{Q\prime}_k \cdot \bold{v}_j^Q)^2 + \sum_{j\in S_{\eta\tau}^Q}(\bold{y} \cdot \bold{v}_j^Q)^2\right\}, 
\end{equation}
where $\left(\{\lambda_k^{C}\}_{k=1}^{N_D},\ \{\bold{v}^{C}_k\}_{k=1}^{N_D}\right)$, $\left(\{\lambda_k^{Q}\}_{k=1}^{N_D}, \{\bold{v}^{Q}_k\}_{k=1}^{N_D}\right)$, 
$\left(\{\lambda_k^{C\prime}\}_{k=1}^{N_D},\ \{\bold{v}^{C\prime}_k\}_{k=1}^{N_D}\right)$, and $\left(\{\lambda_k^{Q\prime}\}_{k=1}^{N_D}, \{\bold{v}^{Q\prime}_k\}_{k=1}^{N_D}\right)$ 
are pairs of the eigenvalues and eigenvectors of $\cov{L}$, $\covq{L}$, $\ntk{L}$, 
and $\ntkq{L}$, respectively. 
Also, $S_{\eta\tau}^C$ and $S_{\eta\tau}^Q$ are the sets of indices where 
$\lambda_j^C<1/\eta\tau$ and $\lambda_j^Q<1/\eta\tau$, respectively; we call 
the eigenvectors corresponding to the indices in $S_{\eta\tau}^C$ or $S_{\eta\tau}^Q$ 
as the bottom eigenvectors. 
That is, now the condition \eqref{equation:general-advantage-condition} is 
converted to the condition \eqref{equation:advantage-mse}, which is represented 
in terms of the eigenvectors of the covariance matrices and the NTKs. 
Of particular importance is the second terms in both sides. 
These terms depend only on how well the bottom eigenvectors of $\ntk{L}$ or 
$\ntkq{L}$ align with the label vector $\bold{y}$. 
Therefore, if the bottom eigenvectors of classically intractable QNTK do not align with $\bold{y}$ at all, while that of classical counterparts align with $\bold{y}$, Eq.~\eqref{equation:advantage-mse} is likely to be satisfied, meaning that we may have 
the advantage of using our qcNN model over classical models. 
This discussion also suggests the importance of the structure of dataset to have quantum 
advantage; see Section 7 of Supplemental Materials of Ref.~\cite{huang2021power}. 
In our case, we may even manipulate $\bold{y}$ so that $\sum_{j\in S_{\eta\tau}^C}(\bold{y} \cdot \bold{v}_j^C)^2 \gg \sum_{j\in S_{\eta\tau}^C}(\bold{y} \cdot \bold{v}_j^Q)^2$ for all possible classical models and thereby obtain a dataset advantageous in the qcNN model. 
A comprehensive study is definitely important for clarifying practical datasets 
and corresponding encoders that achieve \eqref{equation:general-advantage-condition}, 
which is left for future work.

\subsubsection*{Note on the quantum kernel method}
The proposed qcNN model has a merit in the sense of computational 
complexity for the training process, compared to the quantum kernel method. 
As shown in \cite{schuld2021supervised}, 
by using the representer theorem \cite{scholkopf2002learning}, 
the quantum kernel method in general is likely to give better solutions than 
the standard (i.e., the data encoding unitary is used just once) variational 
method for searching the solution, in terms of the training error. 
However, the quantum kernel method is poor in scalability, as in the case of 
the classical counterpart; 
that is, $O(N_D^2)$ computation is needed to calculate the quantum kernel. 
To the contrary, our qcNN is exactly the kernel method in the infinite width 
limit of the classical part, and the 
computational complexity to learn the approximator is $O(N_D T)$ with $T$ 
the number of iterations. 
Therefore, as far as the number of iterations satisfies $T \ll N_D$, 
our qcNN model casts as a scalable quantum 
kernel method.

\subsubsection*{
Specific setting where our model outperforms pure quantum or classical models
}

Secondly, we discuss the possible advantage of the proposed qcNN model over some 
other models, for the training error in the following feature prediction problem 
of quantum states. 
That is, we are given the training set $\{ \rho(\bold{x}^a), y_a\}$, where 
$\rho(\bold{x}^a)$ is an unknown quantum state with $\bold{x}^a$ the 
characteristic input label such as temperature and $y_a$ is the output mean 
value of an observable such as the total magnetization; the problem is, based 
on this training set, to construct a predictor of $y$ for a new label $\bold{x}$ 
or equivalently $\rho(\bold{x})$. 
Let us now assume that the proposed model can directly access to $\rho(\bold{x}^a)$; 
then clearly it gives a better approximator to the training dataset and thereby 
a better predictor compared to any classical model that can only use 
$\{ \bold{x}^a, y_a\}$. 
Also, as shown below Theorem~\ref{theorem:ntkq}, our model can represent a nonlinear 
function of the projected quantum kernel and thus presumably approximates the 
training dataset better than any full-quantum model that can also access to 
$\rho(\bold{x}^a)$ yet is limited to produce a linear function 
$y={\rm Tr}[A U(\boldsymbol{\theta})\rho(\bold{x})U^\dagger(\boldsymbol{\theta})]$ 
with an observable $A$. 
These advantage will be actually numerically demonstrated in Section~\ref{section:exp_qdata}. 
Moreover, Ref.~\cite{huang2022provably} proposed a model that makes a random 
measurement on $\rho(\bold{x}^a)$ to generate a classical shadow for approximating 
$\rho(\bold{x}^a)$ and then constructs a function of the shadows to predict $y$ 
for a new input $\rho(\bold{x})$. 
Note that our model constructs an approximator directly using the randomized 
measurement without constructing the classical shadows and thus includes the 
class of systems proposed in \cite{huang2022provably}; 
hence the former can perform better than the latter. 
Importantly, Ref.~\cite{huang2022provably} identifies the class of problems that can 
be efficiently solved by their model; hence, in principle, this class of problems 
can also be solved by our model. 
Lastly, Ref.~\cite{doi:10.1126/science.abn7293} identifies a class of 
similar feature-prediction problems that can be solved via a specific quantum model 
with constant number of training data but via any classical model with an exponential 
number of training data. 
We will be trying to identify the setting that realizes this provable quantum 
advantage in our qcNN framework.

\section{Numerical experiment}
\label{section:numerical-experiment}

The aim of this section is to numerically answer the following three questions:
\begin{itemize}
    \item 
    How fast is the convergence of QNTK, stated in the theorems in the 
    previous section? 
    In other words, how much is the gap between the training dynamics of an actual 
    finite-width qcNN and that of the theoretical infinite-width qcNN?
    
    \item How much does the locality $m$ (i.e., the size 
    of randomization in qcNN for extracting the features of encoded data) 
    affect on the training of qcNN? 
    
    \item Is there any clear merit of using our proposed 
    qcNN over fully-classical or fully-quantum machine learning models? 
\end{itemize}

To examine these problems, we perform the following three types of numerical 
experiments. 
As for the first question, in Sec.~\ref{section:exp_sin_qc} we compare the 
performance of a finite-width qcNN with that of the infinite-width qcNN in 
specific regression and classification problems; in particular, various types 
of quantum data-encoders will be studied. 
We then examine the second question for a specific regression problem, in 
Sec.~\ref{section:exp_HD_class}. 
Finally, in Sec.~\ref{section:exp_qdata}, we compare the performance of a 
finite-width qcNN with a fully-quantum NN (qNN) as well as a fully-classical NN (cNN), 
in special type of regression and classification problems such that the dataset 
is generated through a certain quantum process. 
Throughout our numerical experiments, we use qulacs \cite{suzuki2021qulacs} to 
simulate the quantum circuit.

\subsection{Finite-width qcNN vs infinite-width qcNN}
\label{section:exp_sin_qc}

In this subsection, we compare the performance of an actual finite-width qcNN with 
that of the theoretical infinite-width qcNN, in a regression task and a 
classification task with various types of quantum data-encoders.

\subsubsection{Experimental settings}

\begin{figure}[h]
    \begin{tabular}{cc}
      \begin{minipage}[t]{0.47\hsize}
        \centering
        \includegraphics[keepaspectratio, scale=0.34]{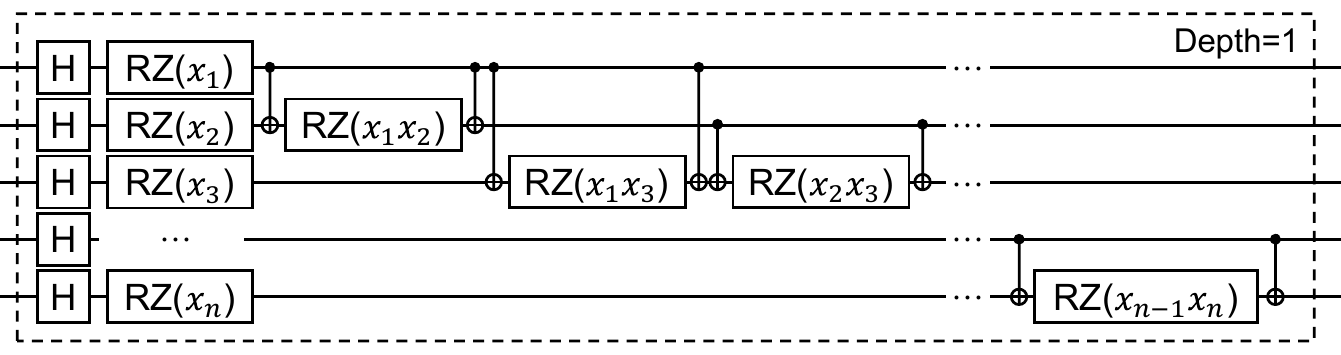}
        \subcaption{Ansatz-A, Ansatz-A4}
        \label{Ansatz_A}
      \end{minipage} &
      \begin{minipage}[t]{0.47\hsize}
        \centering
        \includegraphics[keepaspectratio, scale=0.34]{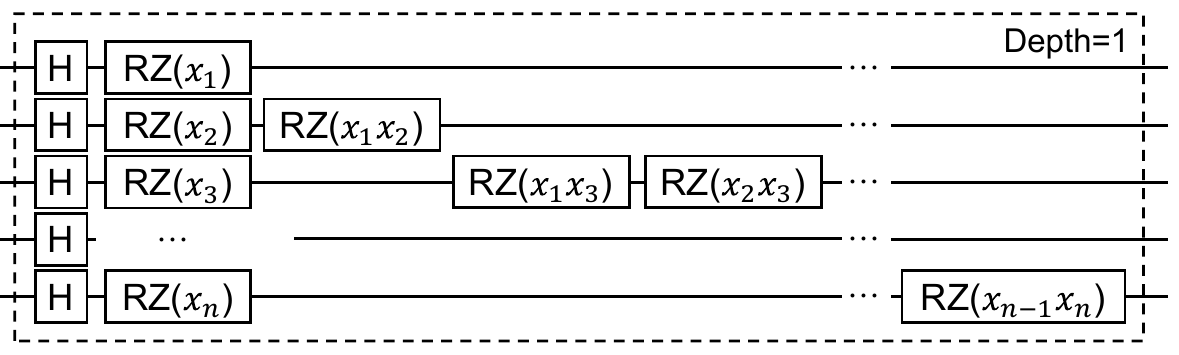}
        \subcaption{Ansatz-A4ne}
        \label{Ansatz_Ac}
      \end{minipage} \\
      \\
      \begin{minipage}[t]{0.47\hsize}
        \centering
        \includegraphics[keepaspectratio, scale=0.34]{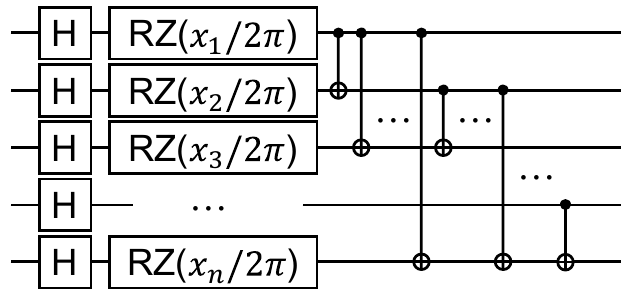}
        \subcaption{Ansatz-B}
        \label{Ansatz_B}
      \end{minipage} &
      \begin{minipage}[t]{0.47\hsize}
        \centering
        \includegraphics[keepaspectratio, scale=0.34]{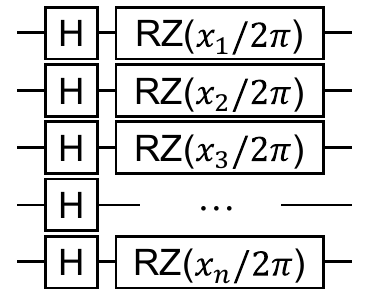}
        \subcaption{Ansatz-Bne}
        \label{Ansatz_Bc}
      \end{minipage} 
    \end{tabular}
     \caption{
     Configuration of $U_{\rm{enc}}(\bold{x})$. 
     First, Hadamard gates are applied to each qubit. 
     Then, the normalized data values $x_i~(i=1,\cdots,n)$ are encoded into 
     the angle of RZ-gates. 
     They are followed by the entangling gate composed of CNOT-gates in (a) and (c). 
     Also, (a) and (b) have RZ-gates whose rotating angles are the product 
     of two data values, which are called as ``Cross-term" in 
     Table~\ref{TABLE_ansatz}. 
     Note that a rotating angle of RZ$(x)$ is $2\pi x$ in (a) and (b), and the 
     dashed rectangle (shown as "Depth=1") is repeated 4 times both in Ansatz-A4 
     and Ansatz-A4ne.}
     \label{FIG_ansatz}
  \end{figure}

{\bf Choices of the quantum circuit.} 
For the quantum data-encoding part, we employ 5 types of quantum circuit 
$U_{\rm{enc}}(\bold{x})$ whose structural properties are listed in 
Table~\ref{TABLE_ansatz} together with Fig.~\ref{FIG_ansatz}. 
In all 5 cases, the circuit is composed of $n$ qubits, and Hadamard gates are 
first applied to each qubit, followed by RZ-gates that encode the data element 
$x_i\in[-1, 1]$ in the form ${\rm RZ}(x_i)={\rm exp}(-2\pi i x_i)$; 
here, the data vector is $\bold{x}=[x_1, x_2, \cdots, x_n]$, meaning that 
the dimension of the data vector is equal to the number of qubits. 
The subsequent quantum circuit is categorized to type-A or type-B as follows. 
As for the type-A encoders, we consider three types of circuits named Ansatz-A, 
Ansatz-A4, and Ansatz-A4ne 
(Ansatz-A4 is constructed via 4 times repetition of 
Ansatz-A); they contain additional data-encoders composed of RZ-gates with 
cross-term of data values, i.e., $x_i x_j$ $(i,j\in [1,2,\cdots,n])$. 
On the other hand, the type-B encoders, Ansatz-B and Ansatz-Bne, which also employ 
RZ gate for encoding the data-variables, do not have such cross-terms, implying 
that the type-A encoders have higher nonlinearity than the type-B encoders. 
Another notable difference between the circuits is the existence of CNOT gates; 
that is, Ansatz-A, Ansatz-A4, and Ansatz-B contain CNOT-gates, while Ansatz-Ane 
and Ansatz-Bne do not (``ne" stands for ``non-entangled").
In general, a large quantum circuit with many CNOT gates may be difficult to 
classically simulate, and thus Ansatz-A, Ansatz-A4, and Ansatz-B are expected 
to show better performance than the other two circuits for some specific tasks. 
The structures of the subsequent classical NN part will be shown in the 
following subsection.

\begin{table}[h]
\begin{center}
\caption{Specific structural properties of $U_{\rm{enc}}(\bold{x})$. }
\label{TABLE_ansatz}
\begin{tabular}{@{}cccccc@{}}
\toprule
\; Circuit type \; & Cross-term \; & CNOT \; & Depth \; \\
\toprule
Ansatz-A   & Yes & Yes & $\times 1$ \\ %
Ansatz-A4  & Yes & Yes & $\times 4$ \\ %
Ansatz-A4ne & Yes &  No & $\times 4$ \\ %
\midrule
Ansatz-B   &  No & Yes & $\times 1$  \\ %
Ansatz-Bne  &  No &  No & $\times 1$  \\ %
\bottomrule
\end{tabular}
\end{center}
\end{table}

{\bf Training method for the classical neural network.} 
In our framework, the trainable parameters are contained only in the classical 
part (cNN), and they are updated via the standard optimization method. 
First, we compute the outputs of the quantum circuit, 
$\fq \xa_i = \langle\psi \xa |U_i^{\dagger} O U_i  |\psi \xa \rangle$,
$i \in [1,2, \ldots, n_0]$, for all the training data 
$\{ (\bold{x}^a, y^a) \},~a \in [1,2,\ldots ,N_D]$; see Fig.~\ref{FIG_model}. 
The outputs are generated through $n_0$ randomized unitaries 
$\{U_1,U_2,\ldots, U_{n_0}\}$, where $U_i$ is sampled from 
unitary 2-designs with the locality $m=1$ \cite{command_haar}. 
We calculate the expectation of $U_i^\dagger O U_i$ directly using the state 
vector simulator instead of sampling (the effect 
of shot noise is analyzed in Sec.~\ref{section:exp_qdata}), and these values 
are forwarded to the inputs to cNN (recall that $n_0$ corresponds to the width 
of the first layer of cNN). 
The training of cNN is performed by using some standard gradient 
descent methods, 
whose type and the hyper-parameters such as the learning rate are appropriately 
selected for each task, as will be described later. 
The parameters at $t=0$ are randomly chosen from the normal distribution 
${\cal N}(0, \sqrt{2 / N_{\rm{param}} })$, where $N_{\rm{param}}$ is the number 
of parameters in each layer (here ${\cal N}(\mu, \sigma)$ is the normal distribution 
with mean $\mu$ and standard deviation $\sigma$).

\subsubsection{Results}
\label{section:exp_sin_reg}

{\bf Result of the regression task.} 
For the regression task, we consider the 1-dimensional hidden function 
$f_{\rm goal}(x)= \rm{sin}(\it{x}) + \epsilon$, where $\epsilon$ is the stochastic 
i.i.d. noise subjected to the normal distribution ${\cal N}(0, 0.05)$. 
The 1-dimensional input data $x$ is embedded into the 4-dimensional vector 
$\bold{x}=[x_1, x_2, x_3, x_4]=[x, x^2, x^3, x^4]$ for quantum circuits.
The training dataset $\{ x^a, f_{\rm goal}(x^a)\}, a=1,\ldots, N_D$ 
is generated by sampling $x\in U(-1, 1)$, where $U(u_1, u_2)$ is the uniform 
distribution in the range $[u_1, u_2]$. 
Here the number of training data point is chosen as $N_D=100$. 
Also the number of qubit is set to $n=4$. 
We use the mean squared error for the cost function and the stochastic gradient 
descent (SGD) with learning rate $10^{-4}$ for the optimizer. 
The cNN has a single hidden-layer (i.e., $L=1$) with the number of nodes 
$n_0 = 10^3$, which is equal to the number of inputs and outputs of cNN.

\begin{figure}[h]
    \centering
    \includegraphics[keepaspectratio, scale=0.5]{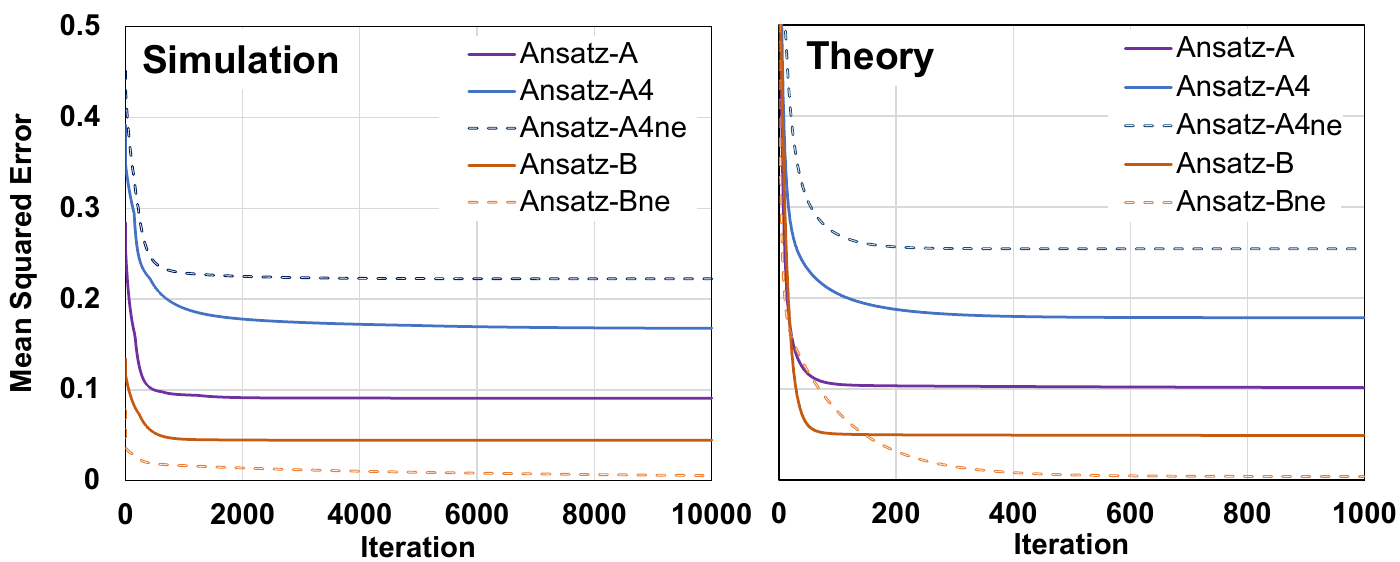}
  \caption{
  Cost function versus the iteration steps for the regression problem. 
  The time-evolution of the cost function obtained by the numerical simulation 
  with $n_0=10^3$ and its theoretical expression assuming $n_0\to\infty$ are 
  shown in the left ``Simulation" and the right ``Theory" figures, respectively.}
  \label{FIG_result_reg}
\end{figure}

The time-evolution of the cost function during the learning process obtained by 
the numerical simulation with $n_0=10^3$ and its theoretical expression assuming 
$n_0\to\infty$ are shown in the left "Simulation" and the right ``Theory" figures, 
respectively, in Fig.~\ref{FIG_result_reg}. 
The curves illustrated in the figures are the best results in total 100 trials 
of choosing $\{U_i\}$ as well as the initial parameters of cNN. 
Notably, the convergent values obtained in the simulation well agree with those of 
theoretical prediction. 
This means that the performance of the proposed qcNN model can be analytically 
investigated for various quantum circuit settings.

Another important fact is that the type-B encoders show better performance than 
the type-A encoders. 
This might be because the type-A encoders have too 
high expressibility for fitting the simple hidden function, which can be 
systematically analysed as demonstrated in \cite{PhysRevA.103.032430, gil2020input}. 
That is, the number of repetition of encoding circuit determines the distribution 
of Fourier coefficients of the model function; if the model function contains 
more frequency components, then it has a bigger expressibility for fitting the 
target function. 
From this perspective, it is reasonable that the type-B encoders (which have only 
single-layer encoding block) show better performance than the type-A encoders 
(which have 4-time-repeating encoding block), since the target hidden function is 
the single-frequency sin function in our setting. 
This observation is actually supported by another result showing that Ansatz-A4 
shows the best performance for a somewhat complicated hidden function 
$f_{\rm goal}(x)= (x-0.2)^2 \rm{sin} (12\it{x})$. 
Summarizing, the encoder largely affects on the overall performance and thus 
should be designed with carefully tuning its expressibility. 
\\

{\bf Result of the classification task.} 
For the classification task, we use an artificial dataset available at \cite{dataset_ad_hoc}, which was used to demonstrate that the quantum support 
vector machine has some advantage over the classical counterpart \cite{havlivcek2019supervised}. 
Each input data vector $\bold{x}$ is of 2 dimensional, and thus the number of 
qubit in the quantum circuit is set as $n=2$. 
The default number of inputs into cNN, or equivalently the width of cNN, is 
chosen as $n_0 = 10^3$; in addition, we will test the cases $n_0 = 10^2$ and 
$n_0 = 10^4$ for the case of Ansatz-A4ne. 
Also, we study two different cases of the number of layers of cNN, as $L= 1$ 
and $L=2$. 
As for the activation function in cNN, we employ the sigmoid function 
$\sigma(q)= 1/(1 + e^{-q})$ for the output layer of both $L=1$ and $L=2$ cases, 
and ReLU $\sigma(q)= {\rm max}(0,q)$ for the input later of the $L=2$ case; 
also the number of nodes is $n_0=10^3$ for the $L=1$ case and $n_0=n_1=10^3$ 
for the $L=2$ case. 
The number of output label $y$ is two, and correspondingly the model yields 
the output label according to the following rule; 
if $f^{C}_{\theta (t)} (f^{Q}(\bold{x}^a))$ is bigger than 0.5, then the output 
label is ``1"; otherwise, the output label is ``0". 
The number of training data is $n_D=50$ for each class. 
As the optimizer for the learning process, Adam \cite{Kingma2014-pa} with 
learning rate $10^{-3}$ is used, and the binary cross entropy 
\eqref{equation:binary-entropy} is employed as the cost function.

\begin{figure}[h]
    \centering
    \includegraphics[keepaspectratio, scale=0.7]{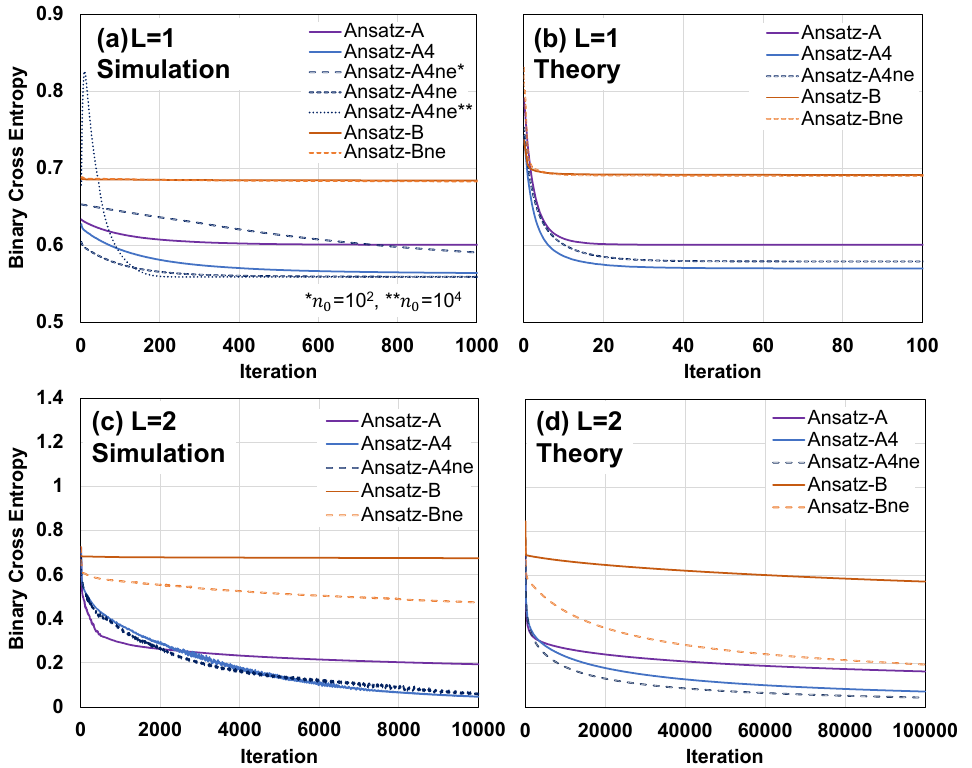}
  \caption{
  Cost function versus the iteration steps for the classification problem. 
  Figures (a, b) and Figures (c, d) depict the results in the case of $L=1$ 
  and $L=2$, respectively. 
  The same dataset is used for each ansatz. }
  \label{FIG_result_class}
\end{figure}

The time-evolution of the cost function during the learning process obtained 
by the numerical simulation and its theoretical prediction corresponding to 
the infinite-width cNN are shown in Fig.~\ref{FIG_result_class}. 
The curves illustrated in the figures are the best results in total 100 trials 
of choosing $\{U_i\}$ as well as the initial parameters of cNN. 
Clearly, the time-evolution trajectories in Simulation and Theory figures for 
the same ansatz are similar, particularly in the case of $L=1$. 
However, there is a notable difference in Ansatz-A4 and Ansatz-A4ne; 
in the Theory figures, the former reaches the final value lower than that achieved 
by the latter, while in the Simulation figures this ordering exchanges. 
Now recall that Ansatz-A4 is the ansatz containing CNOT gates, which induce classically intractable quantum state. 
In this sense, it is interesting that Ansatz-A4 outperforms Ansatz-A4ne, which is though observed only in the case (b) $L=1$ Theory.

In addition, to see the effect of enlarging the width of cNN, we compare three 
cases where the quantum part is fixed to Ansatz-A4ne and the width of cNN varies 
as $n_0 = 10^2, 10^3, 10^4$, in the case of (a) $L=1$ Simulation. 
(Recall that the curve in the Theory figure corresponds to the limit of 
$n_0\to \infty$.) 
The result is that the convergence speed becomes bigger and the value of final 
cost becomes smaller, as $n_0$ becomes larger, which is indeed consistent to the 
NTK theory.

In figures $(c, d)$ for $L=2$, the trajectory from the simulation closely mirrors that of the theory. In particular, the theoretical result successfully predicts which encoder is effective. We also observe that the convergence speed of the theoretical result when $L=2$ is significantly slower than that for $L=1$ due to small eigenvalues in the QNTK. Consequently, the training using a finite-width DNN does not converge within our 10,000-iteration experiment. This results in a large discrepancy between the final cost values in Simulation and Theory in the cases of type-B. In the long iteration limit, we anticipate that the final cost values of both the Simulation and Theory will almost align. Moreover, although the trajectories of type-A from Simulation reaches lower values in fewer iterations than in Theory, this does not necessarily imply that the convergence speed of the Simulation is faster. Even if the convergence speed is not faster than that in Theory, the cost values may still reach smaller values with small steps if the final cost values in the convergence in Simulation are smaller than those in Theory. To examine these properties in convergence, simulation with longer steps are required, which will be addressed in future research.

\begin{figure}[h]
    \centering
    \includegraphics[keepaspectratio, scale=0.8]{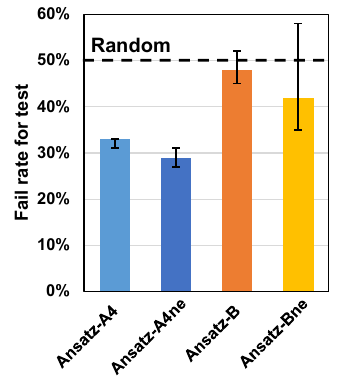}
  \caption{
  Failure rate for the test data ($L=1$, $n_0 = 10^3$). 
  Colored bars represent the median, and the lower (upper) edge of the error bar 
  represents the best (worst) score in total 100 trials. 
  Each scores are calculated with 100 test data. 
  The dashed horizontal line shows the score via the random guess, which is 
  50$\%$ since this is a 2-class classification task. 
  }
  \label{FIG_result_test}
\end{figure}

Finally, to see the generalization error, we input 100 test dataset for the trained 
qcNN models. 
Figure~\ref{FIG_result_test} shows the failure rate, which can be regarded as the 
generalization error, for some types of ansatz. 
Because the failure rate obtained when using the 
classical kernel method presented in \cite{qiskit_qsvm} is $45\%$, Ansatz-A4 
and -A4ne achieve better performance. 
This indicates that qcNN with enough expressibility could have higher performance 
than that of classical method. 
As another important fact, the result is consistent to that of training error; that 
is, the ansatz achieving the lower training error shows the lower test error. 
This might be inconsistent to the following general feature in machine learning; 
that is, too much expressibility leads to the overfitting and eventually degrades 
the machine learning performance. 
However, our model is a function of the projected quantum kernel, which may have 
a good generalization capability as suggested in \cite{huang2021power}. 
Hence our qcNN model achieving small training error would have a good generalization 
capability. 
Further work comparing the performance achieved by full-quantum and full-classical 
methods will be presented in Section.~\ref{section:exp_qdata}.

\subsection{Effect of the locality on the machine learning performance}
\label{section:exp_HD_class}

Here we focus on the locality $m$, i.e., the size of the randomizing unitary gate. 
In our framework, this is regarded as a hyper-parameter, which determines the 
dimension of reduced Hilbert space, $2^m$. 
Note that the system performance may degrade if $m$ is too large as pointed in 
\cite{huang2021power}, and thus $m$ should be carefully chosen. 
Also, $m$ affects on the eigenvalue distribution of QNTK, which closely relates 
to the convergence speed of the learning dynamics. 
Considering the fact that a random circuit may extract essential quantum effect 
in addition to the above-mentioned practical aspect, in this subsection we study 
a specific system and a ML task to analyze how much the locality $m$ affects on 
the convergence speed and the resultant performance. 

The ML task is the classification problem for Heart Disease dataset 
\cite{heart_disease_dataset}. 
This dataset has 12 features, meaning that we use a 12-qubits system to encode 
one feature into one qubit. 
The goal is to use the training dataset to construct a model system that 
predicts if a patient would have a heart disease. 
The number of training data is 100, half of which are the data of patients 
having a heart disease. 
We take the qcNN model with $L=1$ and several values of $m$; 
in particular, we examine the cases $m=1,2,3,4,6$ for the same dataset. 
The other setup including the cost function are the same as that used in the 
previous classification experiment discussed in Sec.~\ref{section:exp_sin_qc}. 

\begin{figure}[h]
    \centering
    \includegraphics[keepaspectratio, scale=0.65]{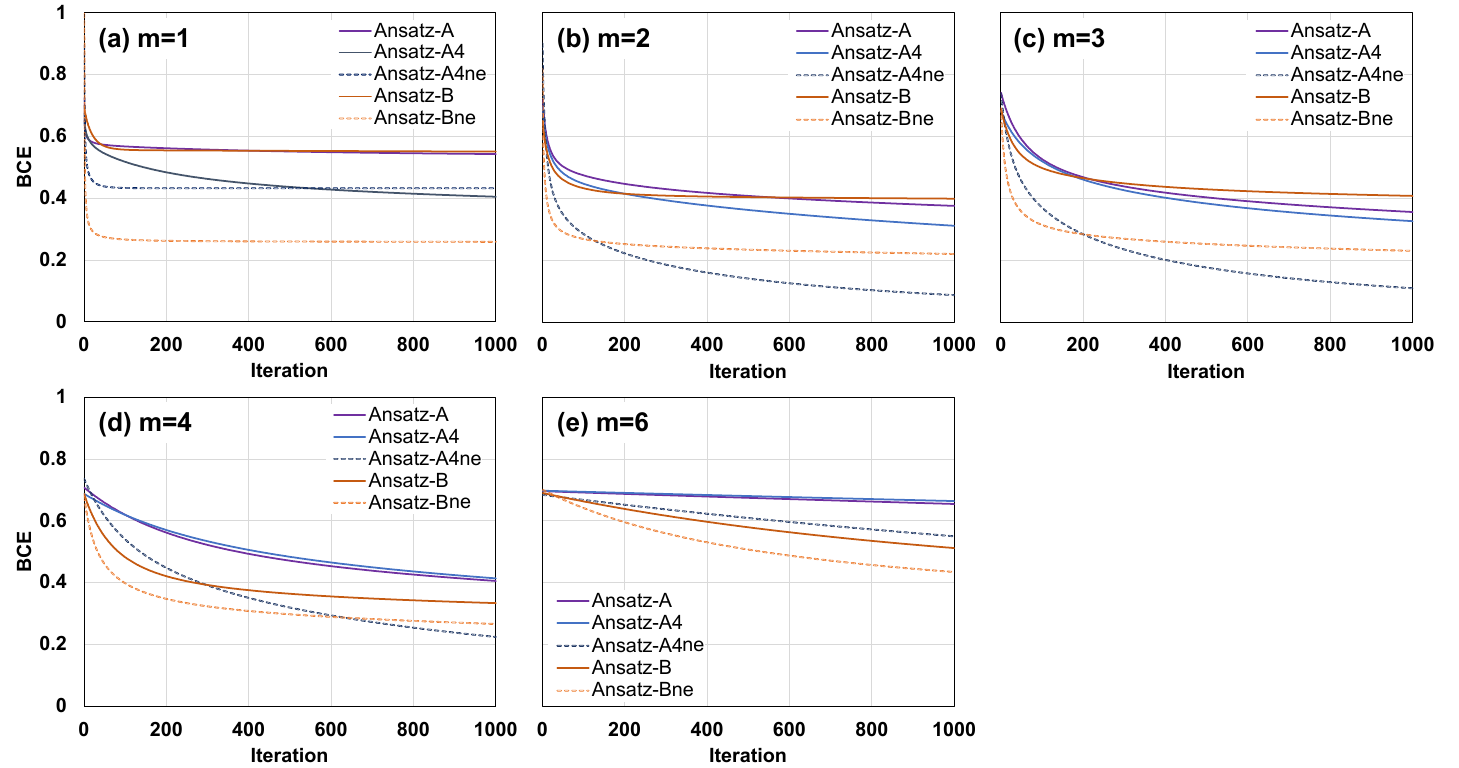}
  \caption{
  Theoretical prediction of learning curve of qcNN for the classification task 
  with Heart Disease Data Set. 
  Figures (a)-(e) correspond to different locality $m$. 
  The quantum circuit is composed of 12 qubits. 
  We use the classical NN with $L=1$. 
  The other setting, including the cost (the binary cross entropy), is the 
  same as that studied in the classification task in Section IV A. 
  }
  \label{FIG_HD_theory}
\end{figure}

We use the theoretical expression of the training process given in 
Eq.~\eqref{equation:theoretical-solution-q}, which is obtained for the 
infinite-width qcNN, rather than simulating the cost via performing actual 
training. 
The learning curves are shown in Fig.~\ref{FIG_HD_theory}. 
As expected, the convergence speed and the value of final cost largely change 
dependent on $m$. 
To understand the mechanism of this result, firstly, let us recall that the 
training curve is characterized by the eigenvalues of QNTK 
$\Theta_Q^{(1)}(\bold{x}, \bold{x}')$ where $\bold{x}$ is the data vector. 
More precisely, as explicitly shown in Eq.~\eqref{equation:theoretical-solution-q}, 
the dynamical component of the index $j$ with large eigenvalue $\lambda_j$ 
converges rapidly, while the component with small eigenvalue does slowly. 
As a result, the distribution of eigenvalues of QNTK determines the entire 
convergence property of training dynamics. 
In particular, the ratio of small eigenvalues is a key to characterize the 
convergence speed. 
In our simulation, we observe that the magnitude of the eigenvalues totally gets 
smaller with larger $m$; this implies that the entire convergence speed would 
decrease, and actually Fig.~\ref{FIG_HD_theory} shows this trend. 
On the other hand, the variance of the eigenvalues distribution also gets smaller 
with larger $m$; as a result, the minimum eigenvalue when $m=2$ is larger than 
that when $m=1$, implying that the training dynamics with $m=2$ may have totally 
better performance in searching the minimum of the cost, than that with $m=1$. 
Therefore, there should be a trade-off in $m$. 
Actually, Fig.~\ref{FIG_HD_theory} clearly shows that, totally, the case of 
$m=2$ or $m=3$ leads to better performance in training. 
This further suggests us to have a conjecture that, in general, a larger value 
of $m$ may not lead to better performance and there is an appropriate value of 
$m$; 
considering the fact that a large random quantum circuit is difficult to 
classically simulate, this observation implies a limitation of the genuine 
quantum part in the proposed qcNN model. 

We further note that the value of final cost, which determines the prediction 
capability for classifying an unseen data, largely changes depending on the 
type of ansatz. 
It is particularly notable that Ansatz-A4ne or Ansatz-Bne achieves the best score for all $m$. 
These are the ansatz that contain no CNOT-gate, and thus the corresponding quantum states are classically simulatable.
That is, for the Heart Disease dataset, it seems that the genuine quantum 
property, including entanglement, is not effectively used for enhancing the classification performance 
of the qcNN model. 
This fact is consistent to the claim given in \cite{huang2021power}, stating 
that all quantum machine learning systems will not improve the performance 
for some specific dataset. 
In the next subsection, therefore, we will show another learning task with 
special type of dataset, such that the proposed qcNN containing CNOT gates 
has certain advantage.

\subsection{Advantage of qcNN over full-classical and full-quantum models}
\label{section:exp_qdata}

Here we study a regression task and a classification task for the dataset 
generated by a quantum process, to demonstrate possible quantum advantage of 
the proposed qcNN model as discussed in Section~III D. 
Our experimental setting is based on the concept suggesting that a quantum 
machine learning model, which is appropriately constructed with carefully 
taking into account the dataset, may have a good learnability and 
generalization capability over classical means. 
In particular, there are some argument discussing quantum advantage for the 
dataset generated through a quantum process; see for example 
\cite{NEURIPS2021_69adc1e1}. 
We show that our qcNN model has such desirable property and actually shows 
better performance, even with much less parameters and thus smaller training 
cost, compared to a fully classical and a fully quantum means for learning 
the quantum data-generating process.

\subsubsection{Machine learning task and models}
\label{section:exp_qdata_qc}

First, we explain the meaning of a data generated from a quantum process, 
which we simply call a ``quantum data" (the case described here is a concrete 
example of the setting addressed in Section~III~D). 
Typically, a quantum data is the output state of a quantum system driven by 
a Hamiltonian $H(\bold{x}^a)$, i.e., 
$\rho(\mathbf{x}^a) = e^{-iH(\bold{x}^a)}\rho_0 e^{iH(\bold{x}^a)}$, where 
the input $\bold{x}^a$ represents some characteristics of the state such as 
controllable temperature. 
The state $\rho(\mathbf{x}^a)$ further evolves through an unknown quantum 
process including a measurement process. 
Finally, the output $y^a$ is obtained by measuring some observables; thus, 
$y^a$ represents a feature of the process or $\rho(\mathbf{x}^a)$ itself. 
Given such training dataset $\{\mathbf{x}^a, y^a\}_{a=1}^{N_D}$, the task is 
to construct a function that approximates this input-output mapping with 
good generalization capability for an unseen input data. 
This problem is related to the general quantum phase recognition (QPR) problem 
\cite{haldane1983nonlinear, pollmann2012detection, cong2019quantum} in 
condensed-matter physics~\cite{sachdev1999quantum}, which is inherently 
classically hard but some quantum machine learning methods may solve 
efficiently \cite{Wu2023quantumphase, okada2022identification, herrmann2022realizing}. 

\begin{figure}[h]
    \begin{tabular}{cc}
    \begin{minipage}[t]{0.47\hsize}
        \centering
        \includegraphics[keepaspectratio, scale=0.6]{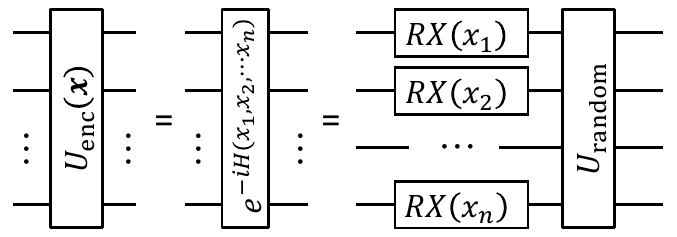}
        \subcaption{Encoder}
        \label{net_Hamiltonian_enc}
    \end{minipage} &
      \begin{minipage}[t]{0.47\hsize}
        \centering
        \includegraphics[keepaspectratio, scale=0.6]{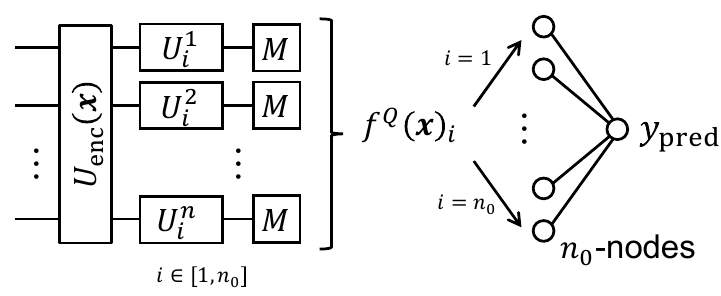}
        \subcaption{qcNN}
        \label{net_qcNN}
      \end{minipage} \\
      \begin{minipage}[t]{0.47\hsize}
        \centering
        \includegraphics[keepaspectratio, scale=0.6]{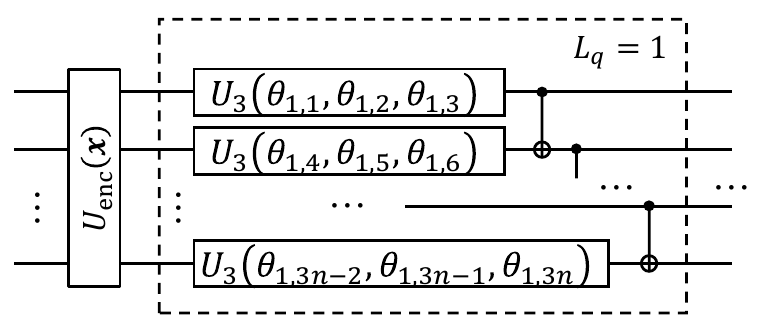}
        \subcaption{qNN}
        \label{net_qNN}
      \end{minipage} &
      \begin{minipage}[t]{0.47\hsize}
        \centering
        \includegraphics[keepaspectratio, scale=0.6]{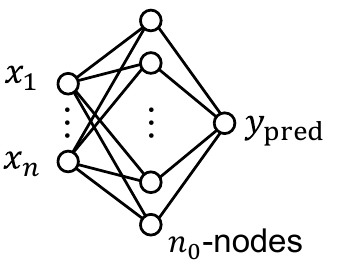}
        \subcaption{pure classical neural network}
        \label{net_cNN}
      \end{minipage} 
    \end{tabular}
     \caption{
     The models used in Sec.~\ref{section:exp_qdata}:
     (a) Quantum circuit for generating the quantum dataset, which is used 
     for the simulation purpose. 
     (b) Quantum-classical hybrid neural network (qcNN),
     (c) quantum neural network (qNN),
     and (d) pure classical neural network (cNN). 
     We use $L_q=10$ in the figure (c), i.e. a 10-layer qNN. 
     In the figure, the box $M$ depicts the measurement and $U_3(\alpha, \beta, \gamma)$ depicts the generic single-qubit rotation gate with 3 Euler angles. 
     }
     \label{FIG_model_qdata}
  \end{figure}

Here we study a specialized version of the above-described problem such that 
the training dataset $\{\mathbf{x}^a, y^a\}_{a=1}^{N_D}$ is provided as follows. 
The input dataset $\{\mathbf{x}^a\}_{a=1}^{N_D}$ is simply generated from 
the $n$-dimensional uniform distribution on $[0, 2\pi]^n$. 
Then $\rho (\mathbf{x}^a)$ is generated via an unknown quantum dynamical process 
$U_{\rm enc}(\bold{x})=e^{-iH(\bold{x})}$; in the simulation, we assume that 
this process is given by the quantum circuit shown in 
Fig.~\ref{FIG_model_qdata}~(a) composed of single qubit $RX$-rotation gates 
followed by a random multi-qubit unitary operator $U_{\rm random}$, the detail 
of which is shown in Appendix~\ref{appendix:exp_qdata}. 
The output $y^a$ is determined depending on the task. 
For the regression task, it is given by $y^a = cg(\mathbf{x}^a)+\epsilon^a$, 
where $g(\mathbf{x})=\mathrm{Tr}\left[\rho(\mathbf{x})O\right]$ and 
$\epsilon$ is a Gaussian noise with $\mathrm{Var}\left[\epsilon\right]=10^{-4}$. 
This measurement process may contain some uncertainties, and thus we assume that 
the observable $O$ is unknown for the algorithms. 
Also $c$ is the normalized constant introduced to satisfy 
$\mathrm{Var}\left[g(\mathbf{x})\right]=1$. 
For the classification task, if $g(\mathbf{x}^a) \geq n/2$ then $y^a=1$, 
and otherwise $y^a=0$, where again $g(\mathbf{x})=\mathrm{Tr}\left[\rho(\mathbf{x})O\right]$. 
In the simulation, we take 
$O=\bigotimes_{i=1}^{n} (\sigma_z^{(i)}+\mathbf{1}^{(i)})/2$, where 
$\sigma_z^{(i)}$ is the Pauli $z$ operator and $\mathbf{1}^{(i)}$ is the 
identity operator on the $i$-th qubit. 
The number of training dataset is chosen as $N_D=1000$ for the regression task 
and $N_D=3000$ for the classification task. 
Moreover, we evaluate the generalization capability using $N_{\rm test}=100$ 
test dataset, which is common for both tasks.

We employ three types of learning models, which are shown in 
Fig.~\ref{FIG_model_qdata}. 
First, the figure (b) shows our qcNN model. 
The point is that this model contains the same encoder $U_{\rm enc}(\bold{x})$ as 
that used for generating the training data; 
that is, the qcNN model has a direct access to the quantum data $\rho(\bold{x}^a)$. 
This process is then followed by the random unitary operator given by the 
product of single qubit gate (i.e., $m=1$). 
The output of the quantum circuit is generated by measuring the observable 
$O=\bigotimes_{i=1}^{n} (\sigma_z^{(i)}+\mathbf{1}^{(i)})/2
=\bigotimes_{i=1}^{n} M^{(i)}$. 
Note that this is the same observable as that used for generating the training 
data, which is assumed to be unknown to the algorithm. 
However, the random unitary process before the measurement makes this assumption 
still valid; actually we found that choosing a different observable other than $O$ 
does not largely change the final performance. 
Then the expectation of measurement results is transferred to the input to 
the single-layer (i.e., $L=1$) cNN. 
The activation function of the output node is chosen depending on the task; 
we employ the identity function for the regression task, while the sigmoid 
function $\sigma(q)= 1/(1 + e^{-q})$ for the classification task. 
Finally, the cNN generates $y_{\rm pred}$ as the raw output for the 
regression task or $y_{\rm pred}$ as the binarized output of the cNN for 
the classification task; that is, as for the latter, $y_{\rm pred}=1$ if the 
output of cNN is bigger than the threshold $0.5$ and $y_{\rm pred}=0$ if the 
output is below $0.5$. 
Note that Ref.~\cite{huang2022provably} uses the random 
measurement to generate a classical shadow for approximating $\rho(\bold{x}^a)$ 
and then constructs a classical machine learning model in terms of the 
shadows to predict $y$ for unseen $\rho(\bold{x})$; 
in contrast, our approach construct a machine learning model directly using 
the randomized measurement result without constructing the classical shadows 
and thus has a clear computational advantage.

The second model is the quantum neural network (qNN) depicted in 
Fig.~\ref{FIG_model_qdata}~(c). 
This model also has the quantum data $\rho(\mathbf{x}^a)$ directly as input, 
as in the qcNN model; that is, in the simulation, the data vector $\bold{x}^a$ 
is encoded to the same quantum circuit $U_{\rm enc}(\bold{x})$ shown in 
Fig.~\ref{FIG_model_qdata}~(a). 
Then a parametrized quantum circuit depicted in the dotted box follows the 
encoder; 
this circuit is repeated $L_q$ times, where each block contains different 
parameters. 
The output of qNN is computed as 
$y_{\rm pred}=w\mathrm{Tr}\left[\rho'(\mathbf{x})O'\right]$, 
where $\rho'(\mathbf{x})$ is the output state of the entire quantum circuit 
and $O'$ is chosen as $O'=\sigma^{(1)}_z$. 
Lastly, $w$ is a scalar parameter; the parameter $w$ is optimized together 
with the circuit parameters $\{\theta_{i,j}\}$, to adjust the output range 
in the case of the regression task, while $w$ is fixed to $1$ in the case 
of the classification task. 
As in the first model, we use $y_{\rm pred}$ for the regression task and its 
binarized version for the classification task.

The third model is the 3-layers cNN depicted in Fig.~\ref{FIG_model_qdata}~(d), 
composed of the $n$-nodes input layer, $n_0$-nodes hidden layer, and a single-node 
output layer. 
The input and the hidden layers are fully connected; the output node is also 
fully connected to the hidden layer. 
We input the data vector $\mathbf{x}^a$ so that its $i$-th component $x_i^a$ is 
the input to the $i$-th node. 
The activation function is chosen depending on the layer and the task; 
we employ the sigmoid function $\sigma(q)= 1/(1 + e^{-q})$ in the hidden 
layer and the identity function in the output node for the regression task, 
while ReLU $\sigma(q)= {\rm max}\{0, q\}$ in the hidden layer and the 
sigmoid function in the output node for the classification task. 
Hence, this pure classical model knows neither the quantum data 
$\rho(\mathbf{x}^a)$ and the observable $O$ (or the model does not 
have enough power for computing these possibly large matrices). 

In all the above three models, we will test four cases in the number of qubit, 
$n=\{2,3,4,5\}$. 
Also, the number of nodes in the cNN is chosen as $n_0=10^3$ for the qcNN model 
and the pure cNN model. 
The expectation value of the observable is calculated using the statevector. 
Adam~\cite{Kingma2014-pa} is used for optimizing the parameters.

\subsubsection{Results}
\label{section:exp_qdata_result}

\begin{figure}[h]
    \centering
    \includegraphics[keepaspectratio, scale=0.5]{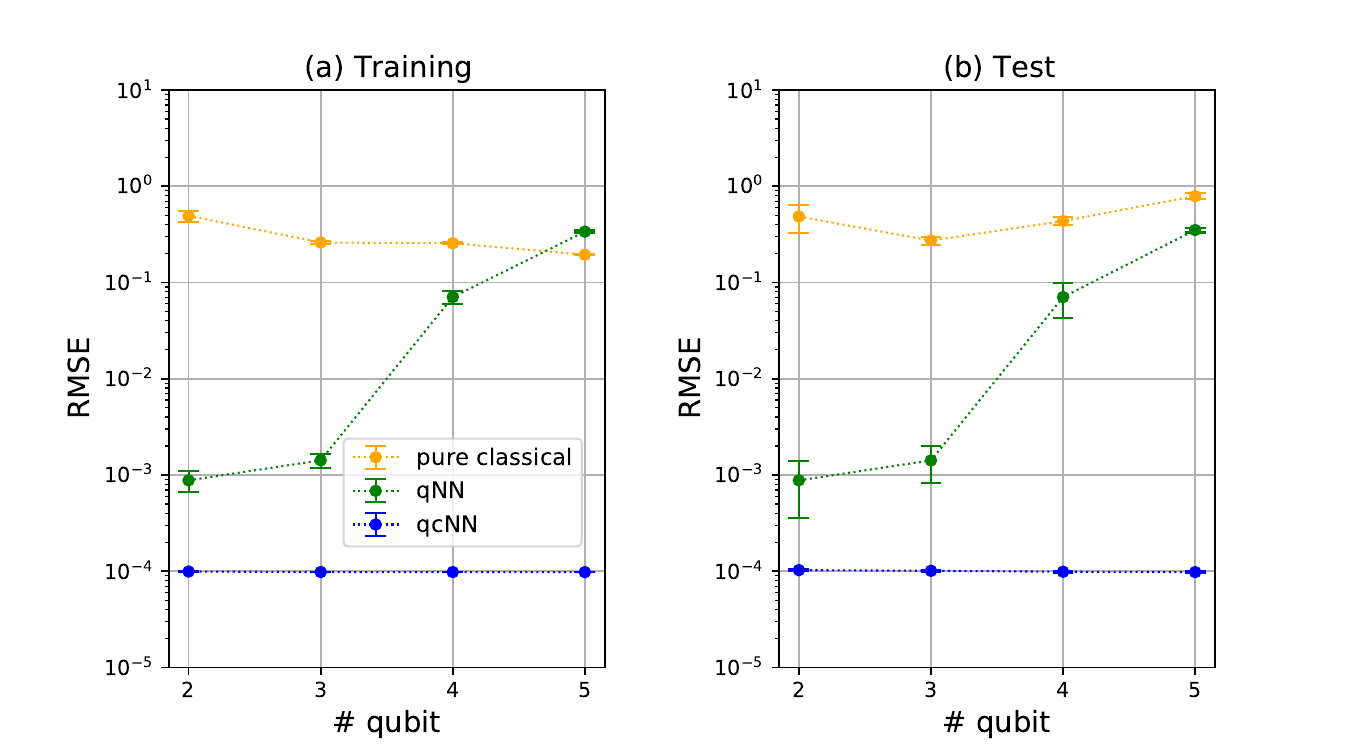}
  \caption{
  The root mean squared errors (RMSE) versus the number of qubits, for 
  (a) the training dataset and (b) the test dataset, for the regression task.}
  \label{FIG_reg_qdata}
\end{figure}

{\bf Result of the regression task.} 
The resulting performance of the regression task, for both training and test 
process, is shown in Fig.~\ref{FIG_reg_qdata}. 
We plot the root mean squared errors (RMSE) between the predicted value 
$y_{\rm pred}$ and the true value, versus the number of qubit $n$. 
For each $n$, we performed five trials of experiments and computed the mean 
and the standard deviation of RMSE. 
Note that the three models have different number of parameters and the measured 
observables (the latter is applied only to the qcNN and qNN models); a detailed 
information is given in Table \ref{TABLE_exp_qdata_param}. %
In particular, the number of parameters of qcNN is much less than that of cNN 
though they have the same width of nodes ($n_0$). 
Also, the number of measured observables required for optimizing qcNN is much 
less than that for qNN, because qcNN model does not need to optimize the quantum 
part by repeatedly measuring the output quantum state.

That is, in total, the qcNN is a compact machine learning model compared to the 
other two. 
Nonetheless, qcNN achieves the best performance for all cases in the number of 
qubits and in both training and test dataset, as shown in Fig.~\ref{FIG_reg_qdata}. 
This is mainly thanks to the ``inductive bias" \cite{NEURIPS2021_69adc1e1}, 
meaning that qcNN model has an inherent advantage of having the quantum data 
itself as input. 
This bias is also given to the qNN model, but this model fails to approximate 
the training and test data when the number of qubits increases; this is 
presumably because the model does not have a sufficient expressibility power 
for representing the target function in the Hilbert space. 
However, even if the qNN model would have such power, it still suffers from 
several difficulties in the learning process such as the barren plateau 
issue and the issue of increasing number of measurement. 
It is also notable that both qcNN and qNN show almost the same performance 
for the training and test dataset, meaning that they do not overfit at all 
to the target data, while the performance of cNN model becomes worse for 
the test data. 
This difference might be because the quantum models can access to the data quantum 
state; this is indeed an inductive bias which may be effectively used for having 
a good generalization capability as suggested in \cite{NEURIPS2021_69adc1e1}.

\begin{figure}[h]
    \centering
    \includegraphics[keepaspectratio, scale=0.5]{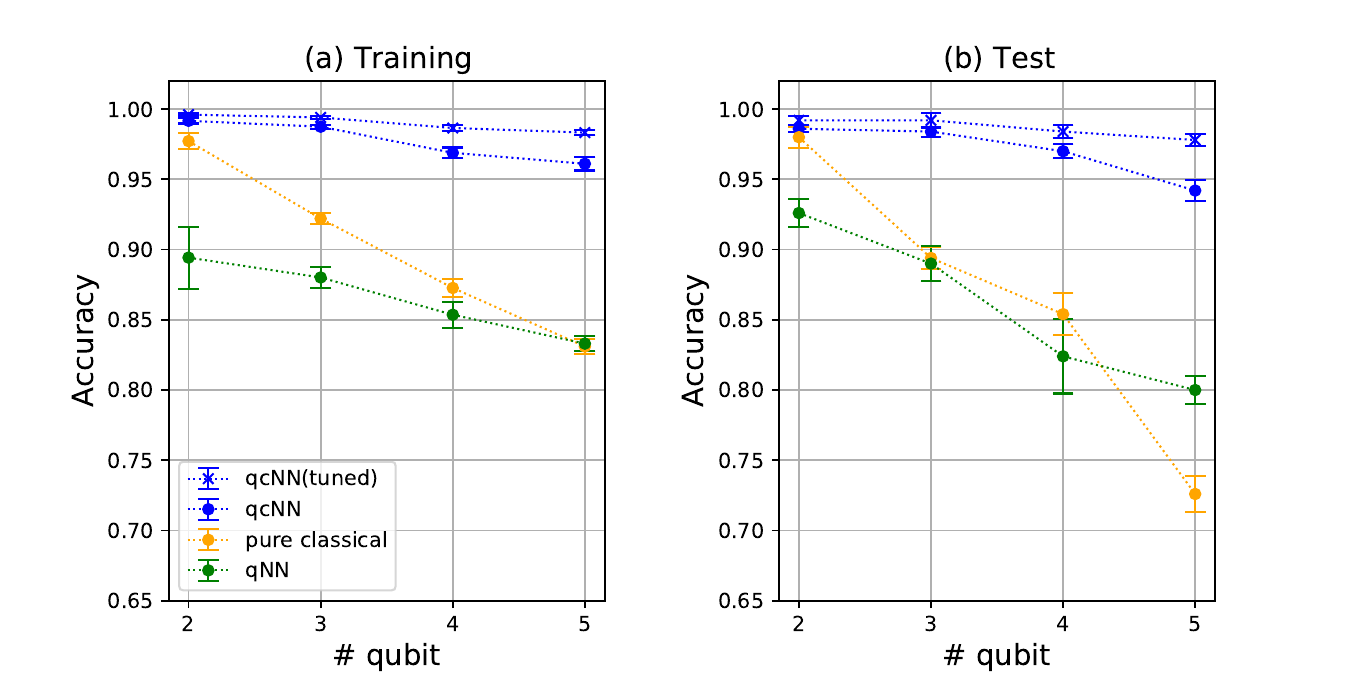}
  \caption{
  The accuracy versus the number of qubits, for (a) the training dataset and 
  (b) the test dataset, for the classification task. 
  }
  \label{FIG_class_qdata}
\end{figure}

{\bf Result of the classification task.} 
The resulting performance of the classification task is shown in 
Fig.~\ref{FIG_class_qdata} which plots the value of Accuracy depending on the number 
of qubits, $n$. 
We perform five trials of experiments and compute the mean and the standard deviation 
of Accuracy. 
In this task, we observed a similar performance trend as that for the regression task, where qcNN shows the best performance for all $n$ presumably due to the same reason 
discussed above; that is, the inductive bias in qcNN model and the lack of 
expressibility of qNN.
In addition to the two class classification task, we also executed the multiclass classification task with the same type of data and models, and observed a similar performance trend (found in Appendix~\ref{appendix:exp_qdata_mclass}).
Just for reference, we also plot Accuracy of qcNN for the case of $N_{\rm ite}=3000$ and 
$n_0=3000$, indicated by `qcNN(tuned)' in Fig.~\ref{FIG_class_qdata} 
(recall that Accuracies of the other three are obtained when $N_{\rm ite}=1000$ 
and $n_0=1000$). 
This means that the performance of qcNN model can be improved via modifying 
purely classical part.

\begin{figure}[h]
    \centering
    \includegraphics[keepaspectratio, scale=0.55]{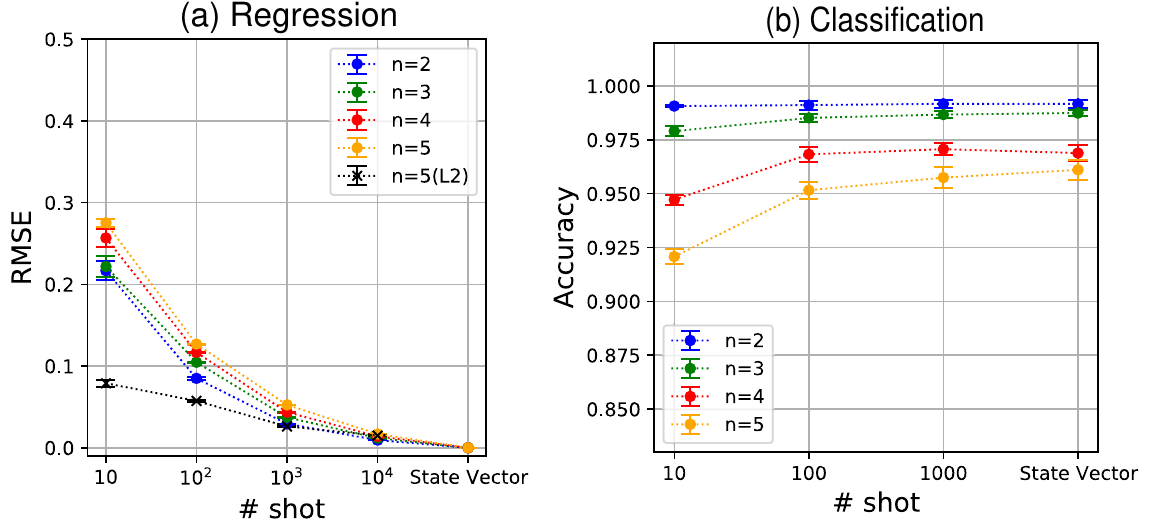}
  \caption{
  (a) The root mean squared errors (RMSE) versus the number of shots for 
  the training dataset, for the regression task. 
  (b) The accuracy versus the number of shots (measurements) for the 
  training dataset, for the classification task.
  }
  \label{FIG_qdata_shot}
\end{figure}

{\bf Effect of shot noise.} 
Finally, we study how the regression/classification performance of the qcNN 
model would change with respect to the number of shots (measurements); 
recall that the previous numerical simulations shown in Figs.~8 and 9 use the 
statevector simulator, meaning that the number of shots is infinite. 
The problem settings, including the type of dataset and the learning model, 
are the same as those studied in the previous subsections. 
The result is summarized in Fig.~\ref{FIG_qdata_shot}, showing (a) the RMSE 
for the regression task and (b) the Accuracy for the classification task. 
In both cases, we examined different number of qubits $n$, where the number 
of layers of cNN part is 1; 
but we also examined the 2-layers cNN only for the case of $n=5$, where the 
width of the 1st and 2nd layers are $n_0=n_1=10^3$. 

In the regression task, we observe a clear statistical trend between RMSE 
$\epsilon$ and the number of shots $N_{\rm shot}$ as 
$\epsilon \sim O(1/\sqrt{N_{\rm shot}})$, except for the case of 2-layers cNN 
denoted as `$n=5$(L2)'. 
The figure suggests that $10^5$ shots seems to offer a sufficient performance 
comparable to the ideal statevector simulator. 
It is notable that the 2-layers cNN significantly reduces RMSE, especially 
when the number of shots is relatively small; 
this seems that the higher nonlinearlity of cNN compensates the shot noise. 
As for the classification task shown in Fig.~\ref{FIG_qdata_shot}~(b), it is 
notable that the necessary number of shots is much smaller compared to the 
regression case. 
In particular, $10^2$ shots achieves a comparable performance to the ideal 
statevector simulator. 
This means that the shot noise, or equivalently the noise contained in the 
input data to the cNN part, does not largely affect on the performance in 
the classification task. 

{\bf Details of the experimental setting in Section IV C.}
The number of model parameters, $N_p$, and the number of different quantum 
circuit required for training, $N_{\rm qc}$, are shown in 
Table~\ref{TABLE_exp_qdata_param}. 
The first and the second row is calculated based on the structure of the models, 
and the third row is the specific $N_p$ in our numerical experiments.
Note that $N_p$ differs depending on the model. 
The value of $N_p$ of qNN seems to be much smaller than others, but this was chosen 
so that the computational cost for training the parameters in the quantum part 
becomes practical (recall that the operation speed of the quantum device is slower 
than that of state-of-the-art classical computers). 
Table~\ref{TABLE_exp_qdata_circ} shows the number of iterations for training, 
$N_{\rm ite}$, together with $N_{\rm qc}$ that is calculated by substituting 
$N_{\rm ite}$ for $N_{\rm qc}$ in Table~\ref{TABLE_exp_qdata_param}.

\begin{center}
\begin{threeparttable}[h]
\caption{The values of model parameters in Sec.~\ref{section:exp_qdata}}
\label{TABLE_exp_qdata_param}
\begin{tabular}{@{}lllccc@{}}
\toprule
\multicolumn{2}{l}{} & ~qcNN~~~ & ~qNN~ & ~cNN \\
\toprule
\# of model parameter\tnote{*} & $N_p$ & ~~~$n_0$ & $3nL_q$ & $(n+2)n_0$ \\
\# of different quantum circuit for training & $N_{\rm qc}$ & ~~~$n_0$ & $2N_p N_{\rm ite}$ & - \\
\midrule
\# of model parameter\tnote{*,**} & $N_p$ & ~~~1k & 60-150 & 4k-7k \\  
\midrule
Final cost value~(Regression, n=5) & $C^{reg}_{f}$ & $9.58 \times 10^{-9}$ & $1.14 \times 10^{-1}$ & $3.78 \times 10^{-2}$ \\
Final cost value~(Classification, n=5) & $C^{class}_{f}$ & 0.251 & 0.503 & 0.364 \\
\bottomrule
\end{tabular}
  \begin{tablenotes}
  \item[*]{Without scaling and bias parameter of the output node.}
  \item[**]{$n_0=1000$, $n=\{2,3,4,5\}$, and $L_q=10$.}
  \end{tablenotes}
\end{threeparttable}
\end{center}

\begin{center}
\begin{threeparttable}[h]
\caption{The number of iterations and different quantum circuits for the 
training process in the problem studied in Sec.~\ref{section:exp_qdata}}
\label{TABLE_exp_qdata_circ}
\begin{tabular}{@{}lllccc@{}}
\toprule
Task $\backslash$ Model & \multicolumn{2}{l}{} & qcNN & qNN & cNN \\
\toprule
\multirow{2}{*}{Regression\tnote{**}} & \# of itetarion    & $N_{\rm ite}$ & 2k-20k & 1k & 2k-20k \\
                                      & \# of different quantum circuit for training & $N_{\rm qc}$ & 1k & 120k-300k & - \\
\hline
\multirow{2}{*}{Classification\tnote{**}~~} & \# of itetarion    & $N_{\rm ite}$ & 1k & 1k & 1k \\
                                            & \# of different quantum circuit for training & $N_{\rm qc}$ & 1k & 120k-300k & -\\
\bottomrule
\end{tabular}
  \begin{tablenotes}
  \item[**]{$n_0=1000$, $n=\{2,3,4,5\}$, and $L_q=10$.}
  \end{tablenotes}
\end{threeparttable}
\end{center}

\section{Conclusion}

In this paper, we studied a qcNN composed of a quantum data-encoder followed 
by a cNN, such that the seminal NTK theory can be directly applied. 
Actually with appropriate random initialization in both parts and by taking the 
large width limit of nodes of the cNN, the QNTK defined for the entire system 
becomes time-invariant and accordingly the dynamics of training process can be 
explicitly analyzed. 
Moreover, we find that the output of the entire qcNN becomes a nonlinear function 
of the projected quantum kernel. 
That is, the proposed qcNN system functions as a nontrivial quantum kernel that 
can processes the regression and classification tasks with less computational 
complexity than that of the conventional quantum kernel method. 
Also, thanks to the analytic expression of the training process, we obtained 
a condition of the dataset such that qcNN may perform better than classical 
counterparts. 
In addition, for the problem of learning the quantum data-generating process, 
we gave a numerical demonstration showing that the qcNN shows a clear advantage 
over full cNNs and qNNs, the latter of which is somewhat nontrivial.

As deduced from the results in Section \ref{section:numerical-experiment} as well 
as those of the existing studies on the quantum kernel method, the performance 
heavily depends on the design of data-encoder and the structure of dataset. 
Hence, given a dataset, the encoder should be carefully designed so that the 
resulting performance would be quantum-enhanced. 
A straightforward approach is to replace the fixed data-encoding quantum part 
with a qNN and train it together with the subsequent data-processing cNN part. 
Actually, we find a general view that a deep learning uses a neural network 
composed of the data-encoding (or feature extraction) part and the subsequent 
data-processing part. 
Hence, such a qNN-cNN hybrid system might have a similar functionality as the 
deep learning, implying that it would lead to better prediction performance and 
further, hopefully, achieve some quantum advantages. 
However, training the qNN part may suffer from the vanishing gradient issue; 
hence relatively a small qNN might be a good choice. 
We leave this problem as a future work.

\begin{acknowledgments}
This work was supported by MEXT Quantum Leap Flagship Program Grant Number JPMXS0118067285 and JPMXS0120319794, and JSPS KAKENHI Grant Number 20H05966. 
\end{acknowledgments}

\bibliographystyle{abb_srt}
\bibliography{main}
\appendix

\section{Proof of Theorem 3}
\label{section:theorem-three}
\begin{customthm}{3}
With $\sigma$ as a Lipschitz function, for $L=1$ and in the limit $n_0\xrightarrow{}\infty$, the output function $\fzero$ is centered Gaussian process whose covaraiance matrix $\covq{1}$ is given by
\begin{equation}
    \covq{1} = \frac{{\rm Tr}(\mathcal{O}^2)}{2^{2m}-1}\sum_{k=1}^{n_Q}\left({\rm Tr}(\rhok{x}\rhok{x^{\prime}}) - \frac{1}{2^m}\right) + \xi^2.
\end{equation}
The reduced density matrix $\rho_x^k$ is defined by
\begin{equation}
\label{equation:projected-density-matrix}
    \rhok{x} = {\rm Tr}_k\left(\uenc(\boldx)|0\rangle^{\otimes n}\langle 0|^{\otimes n}\uenc(\boldx)^{\dagger}\right)
\end{equation}
where ${\rm Tr}_k$ is the partial trace over the Hilbert space associated with all qubits except $(k-1)m \sim km-1$-th qubits.
\end{customthm}
\begin{proof}
From \eqref{equation:quantum-network-propagation} with $L=1$, the prediction function becomes 
\begin{equation}
    \ft(\bold{x}) = \frac{1}{\sqrt{n_0}}W^{(0)}\bold{f}^Q(\bold{x}) + \xi b^{(0)}.
\end{equation}
The distribution of $\fzero$ conditioned on the values of $\bold{f}^Q(\bold{x})$ is centered Gaussian with covariance \begin{equation}
    \label{equation:cov-finite}
    \begin{split}
    Cov^{(1)}(\bold{x}, \bold{x}^{\prime}) &= \frac{1}{n_0}\bold{f}^Q(\bold{x}) \cdot \bold{f}^Q(\bold{x}^{\prime}) + \xi^2 \\
    &= \frac{1}{n_0}\sum_{i=1}^{n_0}\langle \psi(\bold{x})|U_i^{\dagger}O U_i|\psi(\bold{x})\rangle \langle \psi(\bold{x}^{\prime})|U_i^{\dagger}O U_i|\psi(\bold{x}^{\prime})\rangle+ \xi^2,
    \end{split}
\end{equation}
which can be easily shown by using
\begin{equation}
\begin{split}
    \langle W_{ij}^{(0)} \rangle &= 0, \qquad \langle W_{ij}^{(0)}W_{k\ell}^{(0)} \rangle = \delta_{ik}\delta_{j\ell} \\
    \langle b^{(0)}_j \rangle &= 0, \qquad \langle b^{(0)}_j b^{(0)}_k \rangle = \delta_{jk}.
\end{split}
\end{equation}
In the limit $n_0 \rightarrow \infty$, from the weak law of large numbers, 
\begin{equation}
\begin{split}
    Cov(\bold{x}, \bold{x}^{\prime})^{(1)} \rightarrow Cov^{(1)}_\infty(\bold{x}, \bold{x}^{\prime}) &= \int d\mu(U) \langle \psi(\bold{x})|U^{\dagger}OU|\psi(\bold{x})\rangle \langle \psi(\bold{x}^{\prime})|U^{\dagger}OU|\psi(\bold{x}^{\prime})\rangle + \xi^2 \\
    &= \inttwodesign{U_1}\inttwodesign{U_2}\cdots\inttwodesign{U_{n_Q}} \\&\qquad\sum_{k=1}^{n_Q}\langle \psi(\bold{x})|I_{(k-1)m}\otimes U_k^{\dagger}\mathcal{O}U_k\otimes I_{(n_Q-k)m}|\psi(\bold{x})\rangle \\
    &\qquad \times  \sum_{r=1}^{n_Q}\langle \psi(\bold{x}^{\prime})|I_{(r-1)m}\otimes U_r^{\dagger}\mathcal{O}U_r\otimes I_{(n_Q-r)m}|\psi(\bold{x}^{\prime})\rangle + \xi^2,
    \end{split}
\end{equation}
where $\mu(U)$ is the distribution of the random unitary matrix and $\inttwodesign{U_k}$ denotes the integral over unitary 2-designs. By setting  $Q^k(\bold{x})$ to
\begin{equation}
   Q^k(\bold{x}) = \sum_{k=1}^{n_Q}\langle \psi(\bold{x})|I_{(k-1)m}\otimes U_k^{\dagger}\mathcal{O}U_k\otimes I_{(n_Q-k)m}|\psi(\bold{x})\rangle,
\end{equation}
we obtain
\begin{equation}
\label{equation:covariance-sum}
    Cov^{(1)}_\infty(\bold{x}, \bold{x}^{\prime}) = \sum_{k\neq r}\inttwodesign{U^k}Q^k(\bold{x})\inttwodesign{U^r} Q^r(\bold{x}^{\prime}) + \sum_{k=1}^{n_Q}\inttwodesign{U^k}Q^k(\bold{x})Q^k(\bold{x}^{\prime}) + \xi^2.
\end{equation}
The summands of the first and the second terms in \eqref{equation:covariance-sum} can be computed by using the element-wise integration formula for unitary 2-designs \cite{Puchala2017}:
\begin{align}
	\label{equation:two-unitaries}	
	\inttwodesign{U} dU U_{ab} U^{\ast}_{cd} &= \frac{\delta_{ab}\delta_{cd}}{N} \\
		\label{equation:four-unitaries}	
	\inttwodesign{U} U_{ab} U^{\ast}_{cd}U_{ef} U^{\ast}_{gh} &= \frac{1}{N^2-1}\left(\delta_{ac}\delta_{bd}\delta_{eg}\delta_{fh}+\delta_{ag}\delta_{bh}\delta_{ce}\delta_{df} \right) \nonumber\\ 
		&-\frac{1}{N(N^2-1)}(\delta_{ac}\delta_{bh}\delta_{eg}\delta_{fd} + \delta_{ah}\delta_{bd}\delta_{ec}\delta_{fh}), 
\end{align}
where $N$ is the dimension of the unitary matrix. 

For the summand of the first term in \eqref{equation:covariance-sum}, we use \eqref{equation:two-unitaries} and obtain 
\begin{equation}
    \inttwodesign{U_k}  [U_{k}^{\dagger}\mathcal{O}U_k]_{ab} = \inttwodesign{U_k} \sum_{cd}[U_k^{\ast}]_{ca}\mathcal{O}_{cd}[U_k]_{db}
    = \sum_{cd}\delta_{ab}\delta_{cd}\mathcal{O}(\bold{x})_{cd} = \delta_{ab}{\rm Tr}(\mathcal{O}) = 0,
\end{equation}
where in the last equality we use that $\mathcal{O}$ is a traceless operator. Therefore the first term in \eqref{equation:covariance-sum} is zero. 

The summand of the second term in \eqref{equation:covariance-sum} can be written as
\begin{align}
    \inttwodesign{U_k} Q^k(\bold{x})Q^k(\bold{x}^{\prime}) &= \inttwodesign{U_k}{\rm Tr}(U_k^{\dagger}\mathcal{O}U_k\rho^k_{\bold{x}}){\rm Tr}(U_k^{\dagger}\mathcal{O}U_k\rho^k_{\bold{x}^{\prime}})\nonumber\\
    \label{equation:cov-second-summand}
    &= \inttwodesign{U_k}\sum_{a_1b_1}\sum_{a_2b_2}[U_k^{\dagger}\mathcal{O}U_k]_{a_1b_1}[\rho^k_{\bold{x}}]_{b_1a_1}[U_k^{\dagger}\mathcal{O}U_k]_{a_2b_2}[\rho^k_{\bold{x}}]_{b_2a_2},
\end{align}
where $\rho_{\bold{x}}^k$ is defined in \eqref{equation:projected-density-matrix}.
By using \eqref{equation:covariance-sum} the integration of the matrix element can be computed as
\begin{equation}
\label{equation:cov-matrix-element}
\begin{split}
    &\inttwodesign{U_k}[U_k^{\dagger}\mathcal{O}(\bold{x})U_k]_{a_1 b_1}[U_k^{\dagger}\mathcal{O}(\bold{x}^{\prime})U_k]_{a_2 b_2}\\
    &=
    \inttwodesign{U_k}\sum_{c_1,d_1}\sum_{c_2,d_2}[U_k^{\ast}]_{c_1 a_1}\mathcal{O}_{c_1d_1}[U_k]_{d_1 b_1} [U_k^{\ast}]_{c_2 a_2}\mathcal{O}_{c_2d_2}[U_k]_{d_2b_2} \\
    &= \frac{1}{2^{2m}-1}\sum_{c_1,d_1}\sum_{c_2,d_2}\left[\left(\delta_{c_1d_1}\delta_{a_1b_1}\delta_{c_2d_2}\delta_{a_2b_2}+\delta_{c_1d_2}\delta_{a_1b_2}\delta_{d_1u_2}\delta_{b_1a_2}\right)\right. \\
    &\qquad\qquad\left.-\frac{1}{2^m}\left(\delta_{c_1d_1}\delta_{a_1b_2}\delta_{c_2d_2}\delta_{a_2b_1}+\delta_{c_1d_2}\delta_{a_1b_1}\delta_{c_2d_1}\delta_{a_2b_2}\right)\right]\mathcal{O}_{c_1d_1}\mathcal{O}_{c_2d_2}\\
    &= \frac{1}{2^{2m}-1}\left[({\rm Tr}(\mathcal{O}))^2\delta_{a_1b_1}\delta_{a_2b_2} + {\rm Tr}(\mathcal{O}^2)\delta_{a_1b_2}\delta_{b_1a_2}-\frac{1}{2^m}\left\{({\rm Tr}(\mathcal{O}))^2\delta_{a_1b_2}\delta_{a_2b_1}+{\rm Tr}(\mathcal{O}^2)\delta_{a_1b_1}\delta_{a_2b_2}\right\}\right]\\
    &= \frac{{\rm Tr}(\mathcal{O}^2)}{2^{2m}-1}\left(\delta_{a_1b_2}\delta_{a_2b_1}-\frac{1}{2^m}\delta_{a_1b_1}\delta_{a_2b_2}\right),
    \end{split}
\end{equation}
where in the last equality we use $\mathcal{O}$ is traceless. Substituting the result of \eqref{equation:cov-matrix-element} to \eqref{equation:cov-second-summand}, we obtain
\begin{equation}
    \label{cov-second-result}
    \inttwodesign{U_k} Q^k(\bold{x})Q^k(\bold{x}^{\prime}) = \frac{{\rm Tr}(\mathcal{O}^2)}{2^{2m}-1}\left[{\rm Tr}\left(\rho^k_{\bold{x}}\rho^k_{\bold{x}^{\prime}}\right)-\frac{1}{2^m}\right].
\end{equation}
Substituting zero to the first term  in \eqref{equation:covariance-sum} and \eqref{cov-second-result} to the summand of the second term, we can show that the covariance matrix is equal to $\covq{1}$. 

Since the covariance matrix $\covq{1}$ does not depend on the value of $\bold{f}_Q(\bold{x})$ in the limit of $n_0\rightarrow \infty$, the unconditioned distribution of $\ft$ is equal to the conditioned distribution of $\ft$, namely the centered Gaussian process with the covariance $\covq{1}$ in this limit.
\end{proof}

\section{Proof of Theorem 4}
\label{section:proof-theorem-four}
\begin{customthm}{4}
\label{theorem:covq-all}
With $\sigma$ as a Lipschitz function, for $L (>1)$ and in the limit $n_0,n_1, \cdots n_{L-1}\xrightarrow{} \infty$, $\fzero$ is centered Gaussian process whose covariance $\covq{L}$ is given recursively by
\begin{equation}
\label{equation:quantum-cov}
\begin{split}
 \covq{1} &= \frac{{\rm Tr}(\mathcal{O}^2)}{2^{2m}-1}\sum_{k=1}^{n_Q}\left({\rm Tr}(\rhok{x}\rhok{x^{\prime}}) - \frac{1}{2^m}\right) + \xi^2. \\
  \covq{\ell + 1} &=\mathbf{E}_{h \sim \mathcal{N}\left(0, \Sigma_Q^{(\ell)}\right)}\left[\sigma(h(\bold{x})) \sigma\left(h\left(\bold{x}^{\prime}\right)\right)\right]+\xi^2
\end{split}
\end{equation}
where the expectation value is calculated by averaging over centered Gaussian process with covariance $\Sigma_Q^{(L)}$.
\end{customthm}
\begin{proof}
We prove that $\Tilde{\alpha}^{(\ell)}(\bold{x})_j$ for $j=1,2,\cdots,n_\ell$ are i.i.d centered Gaussian process with the covariance given by the equation \eqref{equation:quantum-cov} in the infinite width limit by induction, which becomes the proof for the theorem.

For $L=1$ we can readily show that the distributions of $\Tilde{\alpha}^{(1)}(\bold{x})_j$ are i.i.d centered Gaussian. Then the value of the covariance can be  derived in the same manner as the proof of Theorem~\ref{theorem:covqone}. 

From the induction hypothesis, $\Tilde{\alpha}^{(\ell)}(\bold{x})_j$ for $j=1,2,\cdots,n_\ell$ are i.i.d centered Gaussian process with the covariance given by the equation \eqref{equation:quantum-cov} in the infinite width limit. The element-wise formula for the forward propagation from $\ell$-th layer to the next layer can be written as
\begin{equation}
    \tilde{\alpha}^{(\ell + 1)}(\bold{x})_j = W^{(\ell+1)}_{jk}\sigma(\tilde{{\alpha}}_k^{(\ell)}(\bold{x})) + b^{(\ell)}.
\end{equation}
By using 
\begin{align}
    \langle W_{jk}^{(\ell)}\rangle = 0, \langle W_{jk}^{(\ell)} W_{j^{\prime}k^{\prime}}^{(\ell)}\rangle = \delta_{jj^{\prime}}\delta_{kk^{\prime}},
\end{align}
it can be readily shown that the distributions of $\tilde{\alpha}^{(\ell + 1)}(\bold{x})_j$ conditioned on the values of $\sigma(\tilde{{\alpha}}_k^{(\ell)}(\bold{x}))_k$ are i.i.d. centered Gaussian process with covariance
\begin{equation}
    Cov^{(\ell+1)}(\bold{x}, \bold{x}^{\prime}) = \frac{1}{n_{\ell}}\sum_k \sigma(\tilde{{\alpha}}_k^{(\ell)}(\bold{x}))\sigma(\tilde{{\alpha}}_k^{(\ell)}(\bold{x}^{\prime})) + \xi^2.
\end{equation}
Since the distributions of $\Tilde{\alpha}^{(\ell)}(\bold{x})_k$ for $k=1,2,\cdots,n_\ell$ are i.i.d, so are the distributions of $\sigma(\Tilde{\alpha}^{(\ell)}(\bold{x})_k)$. Therefore from the weak law of large number, in the limit $n_{\ell}\rightarrow \infty$ the sum is transformed to the expectation value as 
\begin{equation}
    Cov^{(\ell+1)}(\bold{x}, \bold{x}^{\prime})\rightarrow \mathbf{E}_{h \sim \mathcal{N}\left(0, \Sigma_Q^{(\ell)}\right)}\left[\sigma(h(x)) \sigma\left(h\left(x^{\prime}\right)\right)\right]+\xi^2.
\end{equation}
Because the limit of the covariance does not depend on $\sigma(\Tilde{\alpha}^{(\ell)}(\bold{x})_k)$, the unconditioned distribution of  $\tilde{\alpha}^{(\ell + 1)}(\bold{x})_j$ is equal to the conditioned distribution, which concludes the proof.  
\end{proof}

\section{Proof of Theorem 5}
\label{section:proof-five}
\begin{customthm}{5}
With $\sigma$ as a Lipschitz function, in the limit $n_0, n_1, \cdots n_{L-1}\xrightarrow{} \infty$, the quantum neural tangent kernel $K_Q^L(\bold{x}, \bold{x}^{\prime}, t)$ converges to the time independent function $\ntkq{L}$, which is given recursively by
\begin{align}
\label{equation:qntk-infinite}
\begin{split}
    \ntkq{1} &= \covq{1} =  \frac{{\rm Tr}(\mathcal{O}^2)}{2^{2m}-1}\sum_{k=1}^{n_Q}\left({\rm Tr}(\rhok{\bold{x}}\rhok{\bold{x}^{\prime}}) - \frac{1}{2^m}\right) + \xi^2,  \\
    \ntkq{\ell+1} &= \ntkq{\ell}\dot{\boldsymbol{\Sigma}}_Q^{(\ell)}\left(\bold{x}, \bold{x}^{\prime}\right)+\boldsymbol{\Sigma}_Q^{(\ell+1)}\left(\bold{x}, \bold{x}^{\prime}\right)
\end{split}
\end{align}
where $\dot{\boldsymbol{\Sigma}}_Q^{(\ell)}\left(\bold{x}, \bold{x}^{\prime}\right)=\mathbf{E}_{h \sim \mathcal{N}\left(0, \boldsymbol{\Sigma}_Q^{(\ell)}\right)}\left[\dot{\sigma}(h(\bold{x})) \dot{\sigma}\left(h\left(\bold{x}^{\prime}\right)\right)\right]$ and $\dot{\sigma}$ is the derivative of $\sigma$.
\end{customthm}
\begin{proof}
We define the elementwise QNTK as
\begin{equation}
    K_{Qjk}^{(\ell)}(\bold{x}, \bold{x}^{\prime}, t)=\sum_{p=1}^P \frac{\partial\tilde{\alpha}^{(\ell)}(\bold{x})_j}{\partial \theta_p(t)}\frac{\partial\tilde{\alpha}^{(\ell)}(\bold{x})_k}{\partial \theta_p(t)}
\end{equation}
and prove 
\begin{equation}
    \label{equation:theorem-kqjk}
     K_{Qjk}^{(\ell)}(\bold{x}, \bold{x}^{\prime}, t) \rightarrow \ntkqx{\ell}{\bold{x}}{\bold{x}^{\prime}} \delta_{jk}
\end{equation}
in the infinite width limit $n_0,n_1,\cdots,n_{\ell-1} \rightarrow \infty$ by induction. Then by setting $\ell=L$ and $n_\ell=1$ we obtain the proof of the theorem.

For $\ell=1$,
\begin{equation}
    \tilde{\bold{\alpha}}^{(1)} (\bold{x}) = \frac{1}{\sqrt{n_0}}W^{(0)}\boldfqx{\bold{x}} + \xi b^{(0)}.
\end{equation}
Then the elementwise QNTK is computed as
\begin{align}
   K_{Qjk}^{(1)}(\bold{x}, \bold{x}^{\prime}, t) &= \frac{1}{n_0}\sum_{i^\prime j^\prime} \frac{\partial  \tilde{\alpha}_{j}^{(1)} (\bold{x})}{\partial W_{i^\prime j^\prime}^{(0)}}\frac{\partial \tilde{\alpha}_{k}^{(1)} (\bold{x}^{\prime})}{\partial W_{i^\prime j^\prime}^{(0)}}
    + \sum_{i^\prime}\frac{\partial \tilde{\alpha}_{j}^{(1)} (\bold{x})}{\partial b^{(0)}_{i^\prime}} \frac{\partial \tilde{\alpha}_{k}^{(1)} (\bold{x})}{\partial b^{(0)}_{i^\prime}}
     \\
    &= 
      \frac{1}{n_0}\sum_{j^{\prime}}\fq({\bold{x}})_{j^{\prime}}\cdot\fq({\bold{x}})_{j^{\prime}} \delta_{jk} + \xi^2 \delta_{jk}\\
   &\rightarrow \covqx{1}{\bold{x}}{\bold{x}^{\prime}} \qquad (n_0\rightarrow \infty),
\end{align} 
where the last line is derived in the proof in Theorem \ref{theorem:covqone}.
Therefore $K_{Qjk}^{(\ell)}(\bold{x}, \bold{x}^{\prime}, t)\rightarrow \ntkqx{1}{\bold{x}}{\bold{x}^{\prime}}=\covqx{1}{\bold{x}}{\bold{x}^{\prime}}$ is proved for $\ell=1$.

From the induction hypothesis, \eqref{equation:theorem-kqjk} holds until $\ell$-th layer in the infinite width limit $n_0, n_1,\cdots, n_{\ell-1}\rightarrow \infty$. Then by using
\begin{equation}
    \tilde{\boldsymbol{\alpha}}^{(\ell + 1)} (\bold{x}) = \frac{1}{\sqrt{n_\ell}}W^{(\ell)}_{jk}\boldsymbol{\alpha}^{(\ell)}(\bold{x}) + \xi b^{(\ell)}.
\end{equation}
\begin{align}
    K_{Qjk}^{(\ell + 1)}(\bold{x}, \bold{x}^{\prime}, t) &= \sum_{\ell^{\prime}=0}^{\ell}
    \sum_{i^\prime j^\prime} \frac{\partial  \tilde{\alpha}_{j}^{(\ell+1)} (\bold{x})}{\partial W_{i^\prime j^\prime}^{(\ell^{\prime})}}\frac{\partial \tilde{\alpha}_{k}^{(\ell+1)} (\bold{x}^{\prime})}{\partial W_{i^\prime j^\prime}^{(\ell^{\prime})}}
    + \sum_{\ell^{\prime}=0}^{\ell}\sum_{i^\prime}\frac{\partial \tilde{\alpha}_{j}^{(\ell+1)} (\bold{x})}{\partial b^{(\ell^{\prime})}_{i^\prime}} \frac{\partial \tilde{\alpha}_{k}^{(\ell+1)} (\bold{x})}{\partial b^{(\ell^{\prime})}_{i^\prime}} \nonumber\\
    &= \kappa^{(0:\ell-1)}(\bold{x}, \bold{x}^{\prime}, t)_{jk} + \kappa^{(\ell)}(\bold{x}, \bold{x}^{\prime}, t)_{jk},
\end{align}
where 
\begin{align}
    \kappa^{(0:\ell-1)}(\bold{x}, \bold{x}^{\prime}, t)_{jk} &= \sum_{\ell^{\prime}=0}^{\ell-1}
    \sum_{i^\prime j^\prime} \frac{\partial  \tilde{\alpha}_{j}^{(\ell+1)} (\bold{x})}{\partial W_{i^\prime j^\prime}^{(\ell^{\prime})}}\frac{\partial \tilde{\alpha}_{k}^{(\ell+1)} (\bold{x}^{\prime})}{\partial W_{i^\prime j^\prime}^{(\ell^{\prime})}}
    + \sum_{\ell^{\prime}=0}^{\ell-1}\sum_{i^\prime}\frac{\partial \tilde{\alpha}_{j}^{(\ell+1)} (\bold{x})}{\partial b^{(\ell^{\prime})}_{i^\prime}} \frac{\partial \tilde{\alpha}_{k}^{(\ell+1)} (\bold{x})}{\partial b^{(\ell^{\prime})}_{i^\prime}} \\
    \kappa^{(\ell)}(\bold{x}, \bold{x}^{\prime}, t)_{jk} &\equiv
    \sum_{i^\prime j^\prime} \frac{\partial  \tilde{\alpha}_{j}^{(\ell+1)} (\bold{x})}{\partial W_{i^\prime j^\prime}^{(\ell)}}\frac{\partial \tilde{\alpha}_{k}^{(\ell+1)} (\bold{x}^{\prime})}{\partial W_{i^\prime j^\prime}^{(\ell)}}
    + \sum_{i^\prime}\frac{\partial \tilde{\alpha}_{j}^{(\ell+1)} (\bold{x})}{\partial b^{(\ell)}_{i^\prime}} \frac{\partial \tilde{\alpha}_{k}^{(\ell+1)} (\bold{x})}{\partial b^{(\ell)}_{i^\prime}} \nonumber\\
    &= \frac{1}{n_\ell}\sum_{j^{\prime}}\tilde{\alpha}(\bold{x})_{j^{\prime}}^{(\ell)}\tilde{\alpha}(\bold{x}^{\prime})^{(\ell)}_{j^{\prime}}\delta_{jk} + \xi^2 \delta_{jk} = Cov^{(\ell+1)}(\bold{x}, \bold{x}^{\prime})\delta_{jk}.
\end{align}
From the proof of Theorem \ref{theorem:covq-all}, $\kappa^{(\ell)}(\bold{x}, \bold{x}^{\prime}, t)_{jk}\rightarrow \covq{\ell}\delta_{jk}$ in the limit $n_{\ell}\rightarrow \infty$.

By using the chain rule
\begin{equation}
    \frac{\partial\tilde{\alpha}(\bold{x})_{j}^{(\ell+1)}}{\partial \theta_p} 
    = \frac{1}{\sqrt{n_\ell}}\sum_{k=1}^{n_\ell}W_{jk}^{(\ell)}\frac{\partial\tilde{\alpha}_{k}^{(\ell)}(\bold{x})}{\partial \theta_p}\dot{\sigma}(\tilde{\alpha}_{k}^{(\ell)}(\bold{x})),
\end{equation}
the other term $\kappa^{(0:\ell-1)}(\bold{x}, \bold{x}^{\prime}, t)_{jk}$ is rewritten as
\begin{equation}
    \kappa^{(0:\ell-1)}(\bold{x}, \bold{x}^{\prime}, t)_{jk} 
    = \frac{1}{n_{\ell}}\sum_{j^\prime k^\prime}W_{jj^\prime}^{(\ell)}W_{kk^\prime}^{(\ell)} K_Q^{(\ell)}(\bold{x}, \bold{x}^{\prime}, t)_{j^\prime k^\prime}
    \dot{\sigma}(\alpha_{j^\prime}^{(\ell)}(\bold{x}))
    \dot{\sigma}(\alpha_{k^\prime}^{(\ell)}(\bold{x}^\prime)),
\end{equation}
and from the induction hypothesis \eqref{equation:theorem-kqjk}, 
\begin{equation}
     \kappa^{(0:\ell-1)}(\bold{x}, \bold{x}^{\prime}, t)_{jk} 
     \rightarrow \frac{1}{n_{\ell}}\sum_{j^\prime}W_{jj^\prime}^{(\ell)}W_{kj^\prime}^{(\ell)}\ntkq{\ell}
    \dot{\sigma}(\tilde{\alpha}_{j^\prime}^{(\ell)}(\bold{x}))
    \dot{\sigma}(\tilde{\alpha}_{j^\prime}^{(\ell)}(\bold{x}^\prime)), 
\end{equation}
in the limit $n_1, n_2, \cdots, n_{\ell-1}$. In the limit $n_{\ell}\rightarrow \infty$ from the weak law of large number, the sum can be replaced by the expectation value as follows:
\begin{align}
    \frac{1}{n_{\ell}}\sum_{j^\prime}W_{jj^\prime}^{(\ell)}W_{kj^\prime}^{(\ell)}\ntkq{\ell}
    \dot{\sigma}(\tilde{\alpha}_{j^\prime}^{(\ell)}(\bold{x}))
    \dot{\sigma}(\tilde{\alpha}_{j^\prime}^{(\ell)}(\bold{x}^\prime))
    &\rightarrow \langle W_{jj^\prime}^{(\ell)}W_{kj^\prime}^{(\ell)}\rangle
    \ntkq{\ell}\dot{\boldsymbol{\boldsymbol{\Sigma}}}_Q^{(\ell+1)}\left(\bold{x}, \bold{x}^{\prime}\right) \nonumber\\
    &= \delta_{jk} \ntkq{\ell}\dot{\boldsymbol{\boldsymbol{\Sigma}}}_Q^{(\ell+1)}\left(\bold{x}, \bold{x}^{\prime}\right).
\end{align}
Thus we have shown that $K_{Qjk}^{(\ell+1)}(\bold{x}, \bold{x}^{\prime}, t) \rightarrow \ntkqx{\ell+1}{\bold{x}}{\bold{x}^{\prime}} \delta_{jk}$, which conclude the proof.
\end{proof}

\section{QNTK with the ReLU activation}
\label{section:relu-analytic}
If we choose the ReLU activation, $\sigma(q)=\max(0, q)$, we can compute the analytical expression of QNTK for $L>1$ recursively. From the formulae proven in Ref.~\cite{Cho2009KernelMF}, the analytic expressions of $\covq{\ell+1}$ and $\dot{\boldsymbol{\Sigma}}^{(\ell)}_Q(\bold{x}, \bold{x}^{\prime})$ are
\begin{align}
\label{equation:cov-relu-1}
    \covq{\ell+1} &= \frac{1}{2\pi}\covqx{\ell}{\bold{x}}{\bold{x}}
    \covqx{\ell}{\bold{x}^{\prime}}{\bold{x}^{\prime}}
    \left(\sin\theta^{\ell}_{\bold{x}\bold{x}^{\prime}}+(\pi-\theta^{\ell}_{\bold{x}\bold{x}^{\prime}})\cos\theta^{\ell}_{\bold{x}\bold{x}^{\prime}} \right)
    + \xi^2 \\
    \label{equation:cov-relu-2}
    \dot{\boldsymbol{\Sigma}}^{(\ell)}_Q(\bold{x}, \bold{x}^{\prime})
    &= \frac{1}{2\pi}(\pi - \theta^{\ell}_{\bold{x}\bold{x}^{\prime}}),
\end{align}
where 
\begin{equation}
    \theta^{\ell}_{\bold{x}\bold{x}^{\prime}}
    = \arccos\left(\frac{\covq{\ell}}{\sqrt{\covqx{\ell}{\bold{x}}{\bold{x}}\covqx{\ell}{\bold{x}^{\prime}}{\bold{x}^{\prime}}}}\right)
\end{equation}
when the activation is given by $\sigma(q)=\max(0, q)$. From \eqref{equation:cov-relu-1}, $\covq{L}$ is recursively computable. By substituting \eqref{equation:cov-relu-1} and \eqref{equation:cov-relu-2} into the latter equation in \eqref{equation:qntk-infinite-1}, $\ntkq{L}$ is also recursively computable.

\section{Proof of Theorem \ref{theorem:positive-ntkq}}
\label{section:qntk-positive}
\begin{customthm}{6}
For a non-constant Lipschitz function $\sigma$, QNTK $\ntkq{L}$ is positive definite unless there exists $\{c_a\}_{a=1}^{N_D}$ such that (i) $\sum_a c_a\rho_{\bold{x}^a}^k=\bold{0}$\ $(\forall k)$, $\sum_a c_a = 0$, and $c_a \neq 0\ (\exists a)$ or (ii) $\xi=0$, $\sum_a c_a\rho_{\bold{x}^a}^k=I_{m}/2^m$\ $(\forall k)$ and $\sum_a c_a = 1$.
\end{customthm}
\begin{proof}
In the recurrence relation,
\begin{equation}
\label{equation:recurrence-again}
\ntkq{\ell+1} = \ntkq{\ell} \dot{\boldsymbol{\Sigma}}_Q^{(\ell)}\left(\bold{x}, \bold{x}^{\prime}\right)+\boldsymbol{\Sigma}_Q^{(\ell+1)}\left(\bold{x}, \bold{x}^{\prime}\right),
\end{equation}
the product of two positive semi-definite kernel, $\ntkq{\ell} \dot{\boldsymbol{\Sigma}}_Q^{(\ell)}\left(\bold{x}, \bold{x}^{\prime}\right)$ is positive semi-definite. Therefore if the rest term of \eqref{equation:recurrence-again}  $\boldsymbol{\Sigma}_Q^{(\ell+1)}\left(\bold{x}, \bold{x}^{\prime}\right)$ is positive definite, $\ntkq{\ell+1}$ is also positive definite.
The positive definiteness of $\covq{\ell+1}$ can be shown by checking if 
\begin{equation}
    \sum_{a,b} c_a c_b \covqx{\ell+1}{\bold{x}^a}{\bold{x}^b} =  \mathbf{E}_{h \sim \mathcal{N}\left(0, \Sigma_Q^{(\ell)}\right)}\left[\left(\sum_a c_a\sigma(h(\bold{x}^a)) \right)^2\right]+\xi^2\left(\sum_{a}c_a\right)^2
\end{equation}
is non-zero for any $\bold{c}\neq\bold{0}$ $\left(\bold{c}=\{c_a\}_{a=1}^{N_D}\right)$, which holds when $\sum_a c_a \sigma(h(\bold{x^a}))$ is not almost surely zero.
If $\covq{\ell}$ is positive-definite the Gaussian $h(\bold{x})$ is non-degenerate, and therefore $\sum_a c_a \sigma(h(\bold{x^a}))>0$ with finite probability since $\sigma$ is not constant function meaning that $\covq{\ell+1}$ is positive definite. Thus the positive definiteness of $\covq{L}$ $(L\geq 2)$ can be recursively proven if $\covq{1}$ is positive definite. 

Recall that 
\begin{equation}
    \covq{1} =  \frac{{\rm Tr}(\mathcal{O}^2)}{2^{2m}-1}\sum_{k=1}^{n_Q}\left({\rm Tr}(\rho_\bold{x}^{k}\rho_{\bold{x}^{\prime}}^{k}) - \frac{1}{2^m}\right) + \xi^2.
\end{equation}
Then $\covq{1}$ is positive definite if 
\begin{equation}
\label{covq-positive-condition}
    \sum_{k=1}^{n_Q}{\rm Tr}\left(\sum_{a}c_a\rho_\bold{x^a}^{k}\right)^2 + \left(\xi^2-\frac{n_Q}{2^m}\right)\left(\sum_a c_a\right)^2 > 0
\end{equation}
for all $\bold{c}\neq \bold{0}$. 

For $\sum_a c_a=0$.
The left hand side of \eqref{covq-positive-condition} becomes $\sum_k{\rm Tr}\left(\sum_{a}c_a\rho_\bold{x^a}^{k}\right)^2$; it becomes zero if and only if $\sum_a c_a \rho^k_{\bold{x}^a}=0$ for all $k$ because $c_a\rho^k_{\bold{x}^a}$ is Hermitian operators, which corresponds to the condition (i) in the theorem. 

For $\sum_a c_a = \beta \neq 0$, the left hand side is proportional to $\beta^2$, thus we can obtain the general condition that \eqref{covq-positive-condition} is satisfied even if we set $\beta=1$. Let us define $\rho^k\equiv\sum_a c_a \rho_\bold{x^a}^{k}$. Then $\rho_k$ is Hermitian with ${\rm Tr}(\rho^k) = 1$. Therefore, given the eigenvalues of $\rho_k$ as $\{\gamma_i^k\}_{i=1}^{2^m}$, 
\begin{equation}
{\rm Tr}\left(\rho^k\right)^2 = \sum_{i=1}^{2^m} \left(\gamma_i^k\right)^2 \geq 2^m\times\sqrt[2^m]{\prod_{i=1}^{2^m} \left(\gamma_i^k\right)^2},
\end{equation}
where equality is attained
when $\gamma_i^k=1/2^{m}$, meaning that ${\rm Tr}\left(\rho^k\right)^2\geq 1/2^m$ and the equality is satisfied when $\rho^k = I_{m}/2^m$.
Thus by using the equality condition, we see that 
\begin{equation}
    \sum_{k=1}^{n_Q}{\rm Tr}\left(\sum_{a}c_a\rho_\bold{x^a}^{k}\right)^2 + \left(\xi^2-\frac{n_Q}{2^m}\right)\left(\sum_a c_a\right)^2 = \xi^2, 
\end{equation}
if and only if $\sum_a c_a \rho_\bold{x^a}^{k} =I_{m}/2^m$. Therefore \eqref{covq-positive-condition} is satisfied unless $\xi^2=0$ and there exists $\bold{c}$ that satisfies $\sum_a c_a=1$, and $\sum_a c_a \rho_\bold{x^a}^{k} =I_{m}/2^m$, which corresponds to the condition (ii).
Since $\covq{1}$ is positive definite unless condition (i) or condition (ii) is satisfied, so is $\ntkq{L}$ as we show above, which concludes the proof of the theorem.
\end{proof}

\section{Multiclass classification task in Section IV C}
\label{appendix:exp_qdata_mclass}
In this section, we demonstrate a multiclass classification task of quantum data.
The basic problem setting is the same as that presented in Sec.~\ref{section:exp_qdata} and Fig.~\ref{FIG_model}, and the output parts are modified for multiclass classification task.
For qcNN and the pure classical neural network model, the output layer is a 4-node full-connected layer, and the activation function of the output layer is the softmax function $f_i(\mathbf{x})=\frac{e^{x_i}}{\sum _{k=1}^4 e^{x_k}}$, where $\mathbf{x}$ corresponds to the 4-dimensional vector of the output layer. 
The predicted label $i$ is given by the index that achieves the highest value of $f_i$.
For qNN model, the ansatz is the same as the original one, and we assigned the predicted label based on the expectation values of the first two qubits, i.e. ``0" for the value ``00", ``1" for ``01", ``2" for ``10", and ``3" for ``11".
The obtained four expectation values are input into the softmax function, and the label with the highest value of function is the predicted label.
We chose the cross entropy as a loss function, and we use Adam for optimizing the model parameters to minimize the loss function.
The target data is prepared as follows.
The quantum data generating process is the same as that presented in Sec. \ref{section:exp_qdata}, and, in this case, we assigned labels $y^a=0,1,2,3$ to 750 samples each in ascending order of the value $g({\bf x}^a)$ since the total number of samples is $N_D=3000$, where $g({\bf x})={\rm Tr}[\rho({\bf x})O]$.
The test data is 100 randomly generated samples.

\begin{figure}[h]
    \centering
    \includegraphics[keepaspectratio, scale=0.45]{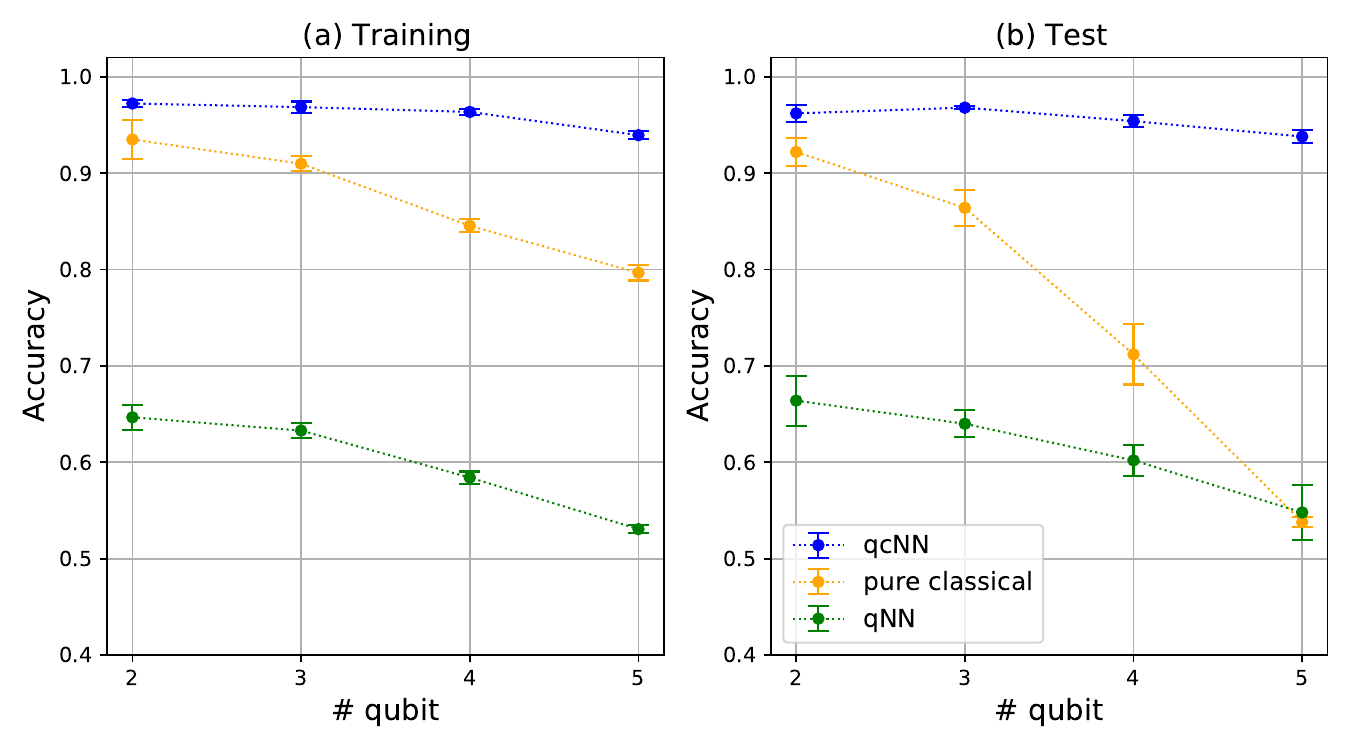}
  \caption{
  The accuracy versus the number of qubits, for (a) the training dataset and 
  (b) the test dataset, for the multiclass classification task. 
  }
  \label{FIG_multi_class_qdata}
\end{figure}

The experimental results are displayed in Fig. \ref{FIG_multi_class_qdata}.
We observed a similar performance trend as that for the two class classification task displayed in Fig.~\ref{FIG_class_qdata}, where qcNN shows the best performance for all $n$ presumably due to the same reason discussed in Sec.~\ref{section:exp_qdata_result}.

\section{Detail of the numerical experiment in Section IV C}
\label{appendix:exp_qdata}
\begin{figure}[h]
    \centering
    \includegraphics[keepaspectratio, scale=0.6]{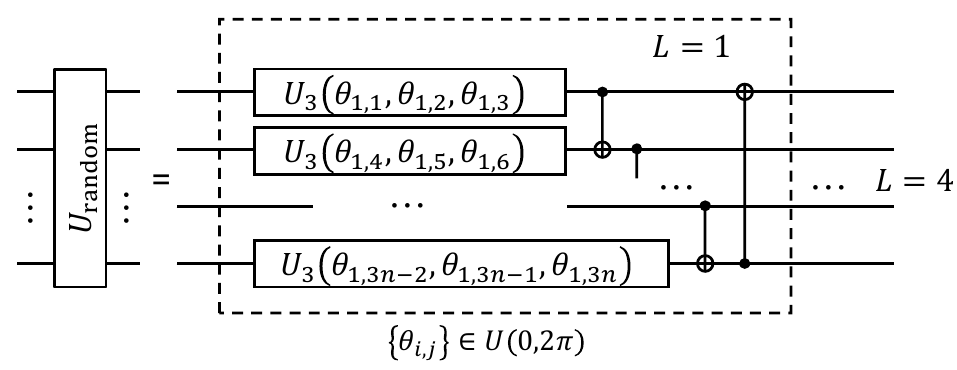}
  \caption{
  The random unitary quantum circuit. $U_3(\alpha, \beta, \gamma)$ depicts the generic single-qubit rotation gate with 3 Euler angles, where $\alpha, \beta, \gamma$ are randomly chosen from the uniform distribution $U(0, 2\pi)$.}
  \label{FIG_gen_random}
\end{figure}

In the numerical experiment, the quasi random unitary circuit shown in Fig.~\ref{FIG_gen_random} is used for $U_{\rm random}$ in Section~\ref{section:exp_qdata}.

\end{document}